\documentclass[twocolumn,preprintnumber,showpacs,aps,epsfig,nofootinbib]{revtex4}

\usepackage{graphicx}
\usepackage{amsfonts}

\usepackage{color}

\usepackage{epstopdf}
\usepackage{latexsym}
\usepackage{amssymb}
\usepackage{amssymb}

\usepackage[center]{subfigure}

\begin{document}

 \newcommand{\bq}{\begin{equation}}
 \newcommand{\eq}{\end{equation}}
 \newcommand{\bqn}{\begin{eqnarray}}
 \newcommand{\eqn}{\end{eqnarray}}
 \newcommand{\nb}{\nonumber}
 \newcommand{\lb}{\label}
\newcommand{\PRL}{Phys. Rev. Lett.}
\newcommand{\PL}{Phys. Lett.}
\newcommand{\PR}{Phys. Rev.}
\newcommand{\CQG}{Class. Quantum Grav.}
\newcommand{\hong}[1]{\textcolor{red}{#1}}

\preprint{IPMU13-0207}

\title{Post-Newtonian approximations in the Ho\v{r}ava-Lifshitz gravity with  extra U(1) symmetry}

\author{Kai Lin ${}^{a,b}$}
\email{lk314159@hotmail.com}

\author{Shinji Mukohyama ${}^{c}$}
\email{shinji.mukohyama@ipmu.jp}

 \author{Anzhong Wang ${}^{a, d}$\footnote{Corresponding author}}
\email{anzhong_wang@baylor.edu}

 \author{Tao Zhu ${}^{a, d}$}
\email{tao_zhu@baylor.edu}

\affiliation{ ${}^{a}$ Institute  for Advanced Physics $\&$ Mathematics,   Zhejiang University of
Technology, Hangzhou 310032,  China \\
${}^{b}$  Instituto de F\'isica, Universidade de S\~ao Paulo, CP 66318, 05315-970, S\~ao Paulo, Brazil \\
${}^{c}$  Kavli Institute for the Physics and Mathematics of the Universe (WPI), TODIAS,  the University of Tokyo,
5-1-5 Kashiwanoha, Kashiwa, 277-8583, Japan \\
${}^{d}$ GCAP-CASPER, Physics Department, Baylor
University, Waco, TX 76798-7316, USA }

\date{\today}

\begin{abstract}
 In this paper, we first propose a universal coupling between the gravity
 and matter in the framework of the Ho\v{r}ava-Lifshitz  theory of
 gravity with an extra U(1) symmetry for both the projectable and
 non-projectable cases. Then, using this universal coupling we study the
 post-Newtonian approximations and obtain the parameterized
 post-Newtonian (PPN) parameters in terms of the coupling constants of
 the theory. Contrary to the previous works in which only two PPN
 parameters were calculated, we obtain {\it all} PPN parameters. We
 then, for the first time in either projectable or non-projectable case, 
 find that all the solar system tests carried out so far are satisfied 
 in a large region of the parameters space. In particular, the same
 results obtained in general relativity can be easily realized here. A
 remarkable feature is that the solar system tests impose no constraint
 on the parameter $\lambda$ appearing in the kinetic part of the
 action. As a result, the solar system tests, when combined with the
 condition for avoidance of strong coupling, do not lead to an upper
 bound on the energy scale $M_{*}$ that suppresses higher dimensional
 operators in the theory. This is in sharp contrast to other versions of
 the HL theory.  [IPMU13-0207] 
\end{abstract}


\maketitle

\section{Introduction}
\renewcommand{\theequation}{1.\arabic{equation}} \setcounter{equation}{0}

In the early of the last century, Physics had experienced two major
revolutionary developments, one was quantum mechanics, which describes
the microscopic world, and the other is  Einstein's general  relativity
(GR), which describes the macroscopic world. When  combining them to
formulate a theory that describes the microscopic  world of gravity, we
have met unprecedented challenges, and so far we are still in a rather
embarrassing situation: such a theory has not been established, and only
a few candidates exist  \cite{QGs}.

Recently, Ho\v{r}ava \cite{Horava} proposed a theory of quantum gravity
in the framework of quantum field theory, in which the
Arnowitt-Deser-Misner (ADM) variables \cite{ADM} are token as the
fundamental ones, with the perspective that Lorentz symmetry appears
only as an emergent symmetry at low energies, but can be fundamentally
absent at high energies. While the breaking of Lorentz symmetry in the
matter sector is highly restricted by experiments, in the gravitational
sector the restrictions are much weaker \cite{LZbreaking,Pola}. The
Lorentz breaking is realized by invoking the anisotropic scaling between
time and space,
\bq
\lb{1.1}
t \rightarrow b^{-z} t,\;\;\; \vec{x} \rightarrow b^{-1}\vec{x}.
\eq
This is a reminiscent of Lifshitz scalars \cite{Lifshitz} in condensed
matter physics, hence the theory is often referred to as the
Ho\v{r}ava-Lifshitz (HL) gravity. Clearly, such a scaling breaks
explicitly the Lorentz symmetry and thus $4$-dimensional diffeomorphism
invariance. Ho\v{r}ava assumed that it is broken only down to
\begin{equation}
\lb{1.2}
 t \to t'(t), \quad \vec{x}\to\vec{x}'(t,\vec{x}),
\end{equation}
the so-called foliation-preserving diffeomorphism,  denoted often by
Diff($M, \; {\cal{F}}$).

Once the general covariance is broken, it immediately results in a
proliferation of independent coupling constants
\cite{Horava,KP,ZWWS}, which could potentially limit the
predictive power of the theory. To reduce the number of independent
coupling constants,  Ho\v{r}ava introduced two independent
conditions, the {\em projectability} and the {\em detailed balance}
\cite{Horava}. The former requires that the lapse function $N$ be a
function of $t$ only,
$N = N(t)$,
while the latter requires that the gravitational potential  should be obtained from a superpotential $W_{g}$,
given by an integral of the gravitational Chern-Simons term ${\omega_{3}(\Gamma)}$ over a 3-dimensional space,
$W_{g}\sim \int_{\Sigma}{\omega_{3}(\Gamma)}$. With these two conditions, the general action contains only five
independent  coupling constants.

The HL theory has soon attracted a lot of attention, and been found
that the projectability condition leads to several undesirable
properties, including infrared instability \cite{Horava,Ins} and strong coupling
\cite{SC,KP,KA,WWa},  although they are not necessarily fundamental
problems~\footnote{It should be noted that the infrared instability does
not show up under a certain condition \cite{Mukohyama:2010xz} and that
the strong coupling is not necessarily a problem if nonlinear effects
help recovering GR at low energy. Of course, the strong coupling implies
that the naive perturbative expansion breaks down and that a proper
non-perturbative treatment is needed.  In fact, in some simplified
situations, fully nonlinear analyses were already performed, and showed
that the $\lambda\to 1$ limit of the theory is continuous and that GR is
recovered in a non-perturbative fashion. Such examples include
spherically symmetric, stationary, vacuum configurations
\cite{Mukohyama:2010xz}, a class of exact cosmological solutions
\cite{WWa} and nonlinear superhorizon perturbations
\cite{Izumi:2011eh,GSWa}. The non-perturbative recovery of GR,
explicitly shown in those examples, may be considered as an analogue of
the Vainshtein effect \cite{Vainshtein:1972sx}.}.
To avoid these issues, various models have been proposed \cite{reviews},
including the healthy extension of the non-projectable HL theory
\cite{BPS}. In the healthy extension, the instability problem was fixed
by the inclusion of the term $a_i a^i$ in the gravitational action,
where $a_i \equiv N_{,i}/N$. The strong coupling problem is resolved by
introducing a new energy scale $M_{*}$, which denotes the suppression
energy of the high-order spatial operators \cite{BPS}. Constraints from
solar system tests and cosmology restrict $M_{*}$ to
$M_{*} \leq 10^{15-16}$ GeV \cite{BPS,BP}.

A dramatical  modification was proposed lately by  Ho\v{r}ava and
Melby-Thompson (HMT) \cite{HMT}, in which an extra local U(1) symmetry
was introduced, so that the symmetry of the theory was enlarged to,
\bq
\lb{symmetry}
 U(1) \ltimes {\mbox{Diff}}(M, \; {\cal{F}}).
\eq
With this symmetry, the spin-0 graviton is eliminated
\cite{HMT,WW}. As a result, all the issues related to it, such as
the infrared instability, strong coupling and different speeds in the
gravitational sector,  are automatically resolved. This was initially
done with the projectability condition and $\lambda = 1$ \cite{HMT},
where $\lambda$, appearing in the kinetic part of the action,
characterizes the IR deviation of the theory from GR. It was soon
generalized to the case with any values of $\lambda$ \cite{daSilva},  in
which the spin-0 gravitons are still eliminated \cite{daSilva,HW}
\footnote{Strictly speaking, in \cite{daSilva,HW} what was shown is that
the spin-0 gravitons are absent in the Minkowski
or (anti-) de Sitter background. A rigorous proof of the absence of spin-0
gravitons can be carried out by analyzing the Hamiltonian structure of
the theory, similar to the projectable case \cite{HMT,Kluson}. In this paper, when we say that the spin-0 gravitons are
eliminated, we always mean that they are absent in terms of linear
perturbations of spacetimes with maximal symmetry, such as the
Minkowski, and (anti-) de Sitter.}.
The consistency of the theory with cosmology was worked out
systematically in \cite{InflationA}. On the other hand, the studies of
solar system tests in the spherical case  showed that the theory is
consistent with observations when the gauge field and the Newtonian
prepotential  are part of the metric \cite{Lin:2012bs}.

A non-trivial generalization of the enlarged symmetry (\ref{symmetry})
to the nonprojectable case $N = N(t, x)$ was also realized  in
\cite{ZWWS},  and has been recently embedded into string theory \cite{JK}. 
In the present paper we shall further generalize the
theory by including two new terms that are compatible with the
symmetry of the theory. We shall denote the coefficients of these new
terms $\sigma_1$ and $\sigma_2$ (see (\ref{2.2})). In the absence of the
new terms, i.e. for $\sigma_1=0=\sigma_2$, it was shown in \cite{ZWWS}
that the only degree of freedom of the theory in the gravitational
sector is the spin-2 massless gravitons, the same as that in
GR. However, as we shall see in the present paper, the scalar graviton
re-emerges in the presence of the new terms \footnote{It is remarkable to 
note that it is exactly because the presence of these terms that the theory becomes
power-counting renromalizable even after the detailed balance condition is
imposed. For detail, see Section III, especially Eqs. (\ref{eq.a1})-(\ref{omegak}) given below. This is
in contrast to all other versions of the HL theory.}. We thus have to revisit
various issues in this version of the HL theory, such as static vacuum
spherical solutions in the IR \cite{LW} and cosmology
\cite{InflationB}. Part of the present paper can be considered as the
first step towards this direction.

In the present paper, we consider the post-Newtonian approximations in
the HL gravity with the enlarged symmetry (\ref{symmetry}) for both the
projectable and non-projectable cases, and derive the parameterized 
post-Newtonian (PPN) parameters in terms of the coupling constants of
the theory. Contrary to the previous works in which only two PPN
parameters were calculated~\cite{Lin:2012bs,LW}, we obtain {\it all} PPN
parameters in both cases. Then, comparing them with the constraints
obtained from the solar system tests carried out so far, we show that
there exists a large region in the parameters space, in which all these
constraints are satisfied. Specifically, the paper is organized as
follows: In Section II, we first give a brief review of the HL theory
with the U(1) symmetry but without the projectability condition. In
doing so, we add a term ${\cal{L}}_{S}$ given by Eq.(\ref{2.2})  into
the action of \cite{ZWWS}, which is also allowed by the enlarged
symmetry (\ref{symmetry}). In fact, with this term, the action takes its
most general form in this version of the HL theory. In Section III, we
study the effects of ${\cal{L}}_{S}$, and find that the spin-0 gravitons
in general appear in the Minkowski background, and are stable. In
Section IV, we consider the coupling of the HL gravity and matter, and
propose a universal coupling. In Section V, using this coupling we study
the post-Newtonian approximations in the case without the projectability
condition, and derive the PPN parameters explicitly in terms of the
coupling constants of the theory. By properly choosing these constants,
the PPN parameters can take the same values as were given  in GR. As a
result, they are consistent with all the solar system tests carried out
so far. The same analysis is presented in Section VI for the projectable
case, and shown that similar conclusion can be reached. In Section VII,
we summarize our main results. Six appendices are also included, in
which various detailed calculations are presented.

\section {Non-Projectable HL Theory with U(1) Symmetry}
\renewcommand{\theequation}{2.\arabic{equation}} \setcounter{equation}{0}

In this section, we shall give a brief review of the  non-projectable HL theory with the enlarged symmetry (\ref{symmetry}), and for detail we refer readers to
\cite{ZWWS}. Another derivation of the general action is presented in Appendix A. The fundamental variables of the theory are
$$
\left(N, \; N_i, \; g_{ij}, \; A, \; \varphi\right),
$$
where $N,\; N_i$ and $g_{ij}$ are, respectively, the
lapse function, shift vector, and 3-metric of the leaves $t = $ Constant  in the ADM decompositions \cite{ADM}.
$A$ and $\varphi$ are, respectively, the gauge field and Newtonian prepotential. Under the local $U(1)$ symmetry,
they   transform as
\bqn
\lb{A.0d}
\delta_{\alpha}A &=&\dot{\alpha} - N^{i}\nabla_{i}\alpha,\;\;\;
\delta_{\alpha}\varphi = - \alpha,\nb\\
\delta_{\alpha}N_{i} &=& N\nabla_{i}\alpha,\;\;\;
\delta_{\alpha}g_{ij} = 0 = \delta_{\alpha}{N},
\eqn
where $N^i \equiv g^{ij} N_j$,  $\alpha$ is   the generator  of the local $U(1)$ gauge symmetry, $\dot{\alpha} \equiv \partial\alpha/\partial t$,
and $\nabla_i$ denotes  the covariant derivative with respect to the 3-metric $g_{ij}$.
Under the Diff($M, \; {\cal{F}}$), they transform as,
\bqn
\lb{A.0e}
\delta{N} &=& \xi^{k}\nabla_{k}N + \dot{N}\xi_0 + N\dot{\xi_0},\nb\\
\delta{N}_{i} &=& N_{k}\nabla_{i}\xi^{k} + \xi^{k}\nabla_{k}N_{i}  + g_{ik}\dot{\xi}^{k}
+ \dot{N}_{i}\xi_0 + N_{i}\dot{\xi_0}, \nb\\
\delta{g}_{ij} &=& \nabla_{i}\xi_{j} + \nabla_{j}\xi_{i} + \xi_0\dot{g}_{ij}, \nb\\
\delta{A} &=& \xi^{i}\nabla_{i}A + \dot{\xi_0}A  + \xi_0\dot{A},\nb\\
\delta \varphi &=& \xi_0 \dot{\varphi} + \xi^{i}\nabla_{i}\varphi,
\eqn
where $\xi^k=\xi^k(x^i,t)$ and $\xi_0=\xi_0(t)$. Note that in this paper we replace $f(t)$ used in
\cite{ZWWS} by $\xi_0(t)$.

The general action reads \cite{ZWWS},
\bqn
\label{action}
S&=&\zeta^2 \int dtd^3x \sqrt{g} N \Big({\cal{L}}_{K}-{\cal{L}}_{V} + {\cal{L}}_{A}
+ {\cal{L}}_{\varphi}\nb\\
&& ~~~~~~~~~~~~~~~~~~~~~~~+{\cal{L}}_S  + {\zeta^{-2}} {\cal{L}}_M\Big),
\eqn
where $g={\rm det}(g_{ij}),\; \zeta^2 \equiv
1/(16\pi G)$ with $G$ being the Newtonian constant of the HL theory, which in principle is different from the Newtonian constant defined
in Eq.(\ref{4.24}).
${\cal{L}}_M$ describes matter fields, and
 \bqn \lb{2.2}
{\cal{L}}_{K} &=& K_{ij}K^{ij} -   \lambda K^{2},\nb\\
{\cal{L}}_{A} &=&\frac{A}{N}\Big(2\Lambda_{g} - R\Big), \nb\\
{\cal{L}}_{\varphi} &=&  \varphi{\cal{G}}^{ij}\big(2K_{ij}+\nabla_i\nabla_j\varphi+a_i\nabla_j\varphi\big)\nb\\
& & +(1-\lambda)\Big[\big(\Delta\varphi+a_i\nabla^i\varphi\big)^2
+2\big(\Delta\varphi+a_i\nabla^i\varphi\big)K\Big]\nb\\
& & +\frac{1}{3}\hat{\cal G}^{ijlk}\Big[4\left(\nabla_{i}\nabla_{j}\varphi\right) a_{(k}\nabla_{l)}\varphi \nb\\
&&  ~~ + 5 \left(a_{(i}\nabla_{j)}\varphi\right) a_{(k}\nabla_{l)}\varphi + 2 \left(\nabla_{(i}\varphi\right)a_{j)(k}\nabla_{l)}\varphi \nb\\
&&~~ + 6K_{ij} a_{(l}\nabla_{k)}\varphi \Big],\nb\\
{\cal{L}}_S &=&\sigma  (\sigma_1 a^ia_i+\sigma_2a^i_{\;\;i}).
 \eqn
Here $\Delta \equiv g^{ij}\nabla_{i}\nabla_{j}, \; \hat{\cal G}^{ijlk} =  g^{il}g^{jk} - g^{ij}g^{kl}$,  and $\Lambda_{g}$ is a coupling constant.
The Ricci scalar  and tensor  are  defined, respectively, as $R = g^{ij}R_{ij}$ and $R_{ij} = g^{kl}R_{kilj}$ in terms of the   Riemann tensor   $R_{i jkl}$,
where
 \bqn \lb{2.6}
 R_{ijkl} &=& g_{ik}R_{jl}   +  g_{jl}R_{ik}  -  g_{jk}R_{il}  -  g_{il}R_{jk}\nb\\
 &&    - \frac{1}{2}\left(g_{ik}g_{jl} - g_{il}g_{jk}\right)R,\nb\\
K_{ij} &\equiv& \frac{1}{2N}\left(- \dot{g}_{ij} + \nabla_{i}N_{j} +
\nabla_{j}N_{i}\right),\nb\\
{\cal{G}}_{ij} &\equiv& R_{ij} - \frac{1}{2}g_{ij}R + \Lambda_{g} g_{ij},\nb\\
a_{i} &\equiv& \frac{N_{,i}}{N},\;\;\; a_{ij} \equiv \nabla_{j} a_{i},\nb\\
{\cal{A}} &\equiv& - \dot{\varphi}  + N^i\nabla_i\varphi
+\frac{1}{2}N\left(\nabla^i\varphi\right)\left(\nabla_i\varphi\right),\nb\\
\sigma &\equiv &\frac{A - {\cal{A}}}{N}.
 \eqn

It should be noted that in writing  the general action (\ref{action}), we have added the term ${\cal{L}}_S$, which is also comparable with the enlarged symmetry (\ref{symmetry}). In fact,
the action (\ref{action})  now represents the most general action for the nonprojectable general covariant theory of the    HL gravity.

When the projectability condition $N = N(t)$  is abandoned, it gives rise to
a proliferation of a large number of independently coupling constants \cite{KP,BPS,ZWWS}. Following Ho\v{r}ava, the detailed balance condition is generalized  to
\cite{ZWWS}\footnote{Note a sign difference in the front of $g_{ij}$ in the $2\times2$ matrix. As shown below, this is required by the stability of the spin-0 gravitons, which are
eliminated  in the version presented in \cite{ZWWS}, in which the term ${\cal{L}}_S$ is absent.},
\bq
\lb{gdbc}
{\cal{L}}_{(V,D)} = \Big(E_{ij} \;\; A_{i}\Big)\left(\matrix{{\cal{G}}^{ijkl} & 0\cr 0 &  g^{ij}\cr}\right)\left(\matrix{E_{kl}\cr A_{j}\cr}\right),
\eq
where ${\cal{G}}^{ijkl}$ denotes the generalized De Witt metric, defined as ${\cal{G}}^{ijkl} = \frac{1}{2} \big(g^{ik}g^{jl} + g^{il}g^{jk}\big) - \lambda g^{ij}g^{kl}$,
and $E_{ij}$ and $A_{i}$ are given by
\bq
\lb{gdbc1}
E^{ij} = \frac{1}{\sqrt{g}}\frac{\delta{W}_{g}}{\delta{g}_{ij}},\;\;\;
A^i = \frac{1}{\sqrt{g}}\frac{\delta W_a}{\delta a_i}.
\eq
The super-potentials $W_g$ and $W_a$ are constructed as \footnote{Note that in \cite{VS} a different generalization was proposed, in which ${\cal{L}}_{(V,D)}$
takes the same form in terms of $E_{ij}$ and ${\cal{G}}^{ijkl}$, as that given in \cite{Horava}, but now the superpotential includes a term $a_ia^i$, i.e.,
$W_{g} = \frac{1}{w^{2}}\int_{\Sigma}{\omega_{3}(\Gamma)} + \int{d^3x \sqrt{g}\left[\mu(R - 2\Lambda) + \beta a_ia^i\right]}$. However, in this generalization, the six-order derivative term
$ \left(\Delta{a^{i}}\right)^{2}$ does not exist in the potential $ {\cal{L}}_{V}$, and is a particular case of Eq.(\ref{2.5a}) with $\beta_8 = 0$. As to be shown below, without this term,
the six-order derivative terms are absent for scalar perturbations. As a result,  the corresponding  theory is not power-counting renormalizable, as already noted in  \cite{ZWWS}.},
\bqn
\lb{gdbc2}
W_{g}  &=& \frac{1}{w^{2}}\int_{\Sigma}{\omega_{3}(\Gamma)} + \mu \int{d^3x \sqrt{g}\Big(R - 2\Lambda\Big)},\nb\\
W_{a} &=& \int{d^{3}x \sqrt{g} \sum_{n= 0}^{1}{{\cal{B}}_{n}a^{i}\Delta^{n}{a_{i}}}},
\eqn
where $\omega_{3}(\Gamma)$ denotes the gravitational 3-dimensional   Chern-Simons term, $w,\; \mu,\; \Lambda$ and ${\cal{B}}_{n}$ are arbitrary  constants. However, to have a healthy
infrared limit,   the detailed balance condition is allowed to be broken softly, by adding all the low dimensional operators, so that the potential finally takes the form
\cite{ZWWS},
\bqn
\lb{2.5a}
 {\cal{L}}_{V} &=&  \gamma_{0}\zeta^{2}  -  \Big(\beta_0  a_{i}a^{i}- \gamma_1R\Big)
+ \frac{1}{\zeta^{2}} \Big(\gamma_{2}R^{2} +  \gamma_{3}  R_{ij}R^{ij}\Big)\nb\\
& & + \frac{1}{\zeta^{2}}\Bigg[\beta_{1} \left(a_{i}a^{i}\right)^{2} + \beta_{2} \left(a^{i}_{\;\;i}\right)^{2}
+ \beta_{3} \left(a_{i}a^{i}\right)a^{j}_{\;\;j} \nb\\
& & + \beta_{4} a^{ij}a_{ij} + \beta_{5}
\left(a_{i}a^{i}\right)R + \beta_{6} a_{i}a_{j}R^{ij} + \beta_{7} Ra^{i}_{\;\;i}\Bigg]\nb\\
& &
 +  \frac{1}{\zeta^{4}}\Bigg[\gamma_{5}C_{ij}C^{ij}  + \beta_{8} \left(\Delta{a^{i}}\right)^{2}\Bigg],
 \eqn
where all the coefficients, $ \beta_{n}$ and $\gamma_{n}$, are
dimensionless and arbitrary, except for the ones of the sixth-order derivative terms,  $\gamma_{5}$ and $\beta_{8}$, which are positive,
$ \gamma_{5} > 0, \; \beta_{8} >  0$, as can be seen from Eqs.(\ref{gdbc})-(\ref{gdbc2}). The coupling constant $\gamma_0$ is related to the cosmological constant
via the relation,
\bq
\lb{2.2d}
\Lambda = \frac{1}{2}\gamma_0 \zeta^2.
\eq
The Cotton tensor $C_{ij}$  is defined as
\bq
\lb{2.2c}
C^{ij} =  \frac{e^{ijk}}{\sqrt{g}} \nabla_{k}\Big(R^{j}_{l} - \frac{1}{4}R\delta^{j}_{l}\Big),
\eq
with  $e^{123} = 1$, etc.   In terms of $R_{ij}$ and $R$, we have \cite{ZWWS},
\bqn
\lb{2.2da}
C_{ij}C^{ij}
&=& \frac{1}{2}R^{3} - \frac{5}{2}RR_{ij}R^{ij} + 3 R^{i}_{j}R^{j}_{k}R^{k}_{i}  +\frac{3}{8}R\Delta R\nb\\
& &  +
\left(\nabla_{i}R_{jk}\right) \left(\nabla^{i}R^{jk}\right) +   \nabla_{k} G^{k},
\eqn
where
\lb{2.2e}
\bqn
G^{k}=\frac{1}{2} R^{jk} \nabla_j R - R_{ij} \nabla^j R^{ik}-\frac{3}{8}R\nabla^k R.
\eqn

Then, the variations of $S$ with respect to $N$ and $N_{i}$ give
rise to the Hamiltonian and momentum constraints, respectively, 
\bqn
\label{hami}
&&  {\cal{L}}_K + {\cal{L}}_V^R + F_V-F_\varphi-F_\lambda+{\cal H}_S = 8 \pi G
 J^t,\\
 \label{mom}
&& M_S^i+\nabla_j \bigg\{\pi^{ij} -(1-\lambda)g^{ij}\big(\nabla^2\varphi+a_k\nabla^k\varphi\big)\nb\\
&& ~~~~~~~~~~~~~~~ - \varphi {\cal{G}}^{ij} - \hat{{\cal{G}}}^{ijkl} a_l \nabla_k \varphi\bigg\} = 8\pi G J^i, ~~~~
 \eqn
where
 \bqn
{\cal{L}}_V^R&=&\gamma_0 \zeta^2+\gamma_1R+\frac{\gamma_2 R^2+\gamma_3 R_{ij}R^{ij}}{\zeta^2}+\frac{\gamma_5}{\zeta^4} C_{ij}C^{ij},\nb\\
{\cal H}_S&=&\frac{2\sigma_1}{N}\nabla_i\left[a^i\left(A-{\cal A}\right)\right]-\frac{\sigma_2}{N}\nabla^2\left(A-{\cal A}\right)\nb\\
&&+\frac{1}{2}\left(\sigma_1a_ia^i+\sigma_2a_i^i\right)\nabla_j\varphi\nabla^j\varphi,\nb\\
M_S^i&=&-\frac{1}{2}\left(\sigma_1a_ka^k+\sigma_2a_k^k\right)\nabla^i\varphi, \nb\\
 J^i &=& -N \frac{\delta {\cal{L}}_M}{\delta N_i},\nb\\
J^t &=& 2 \frac{\delta (N {\cal{L}}_M)}{\delta N},\nb\\
\pi^{ij}&=&-K^{ij}+\lambda K g^{ij},
 \eqn
and $F_V,\;F_\varphi$ and
$F_\lambda$ are given by Eqs.(\ref{a1})-(\ref{a3}) in Appendix B.
Variations of $S$ with respect to $\varphi$ and $A$ yield,
respectively,
\bqn
\label{phi}
&& \frac{1}{2} {\cal{G}}^{ij} ( 2K_{ij} + \nabla_i\nabla_j\varphi  +a_{(i}\nabla_{j)}\varphi)\nb\\
&& + \frac{1}{2N} \bigg\{ {\cal{G}}^{ij} \nabla_j\nabla_i(N \varphi) - {\cal{G}}^{ij} \nabla_j ( N \varphi a_i)\bigg\}\nb\\
&& - \frac{1}{N} \hat{{\cal{G}}}^{ijkl} \bigg \{ \nabla_{(k} ( a_{l)} N K_{ij}) + \frac{2}{3} \nabla_{(k} (a_{l)} N \nabla_i \nabla_j \varphi)\nb\\
&& - \frac{2}{3} \nabla_{(j} \nabla_{i)} (N a_{(l} \nabla_{k)} \varphi) + \frac{5}{3} \nabla_j (N a_i a_k \nabla_l \varphi)\nb\\
&& + \frac{2}{3} \nabla_j (N a_{ik} \nabla_l \varphi)\bigg\}\nb\\
&& -\frac{1}{2N}\Bigg\{\frac{1}{\sqrt{g}}\frac{\partial}{\partial t}\left[\sqrt{g}\left(\sigma_1a_ia^i+\sigma_2a^i_i\right)\right]\nb\\
&&-\nabla_k\left[\left(N^k+N\nabla^k\varphi\right)\left(\sigma_1a_ia^i+\sigma_2a_i^i\right)\right]\Bigg\}\nb\\
&& + \frac{1-\lambda}{N} \bigg\{\nabla^2  \left.[N (\nabla^2 \varphi + a_k \nabla^k \varphi)\right.] \nb\\
&& - \nabla^i [N(\nabla^2 \varphi + a_k \nabla^k \varphi) a_i] \nb\\
&&+ \nabla^2 (N K) - \nabla^i ( N K a_i)\bigg \}
 = 8 \pi G J_\varphi,
\eqn
and
 \bqn
 \label{A}
 R-2 \Lambda_g- \sigma_1a_ia^i-\sigma_2a_i^{\;\; i} = 8 \pi G J_A,
 \eqn
where
 \bqn
J_\varphi = -\frac{\delta {\cal{L}}_M}{\delta \varphi},\;\;\;
 J_A= 2 \frac{\delta ( N {\cal{L}}_M)}{\delta A}.
\eqn
Eqs.(\ref{phi}) and (\ref{A}) will be referred, respectively,  to as $\varphi$- and $A$- constraint.

On the other hand, the variation of $S$ with respect to
$g_{ij}$ yields the dynamical equations,
 \bqn
 \label{dyn}
\frac{1}{\sqrt{g}N} \frac{\partial}{\partial t}\left(\sqrt{g} \pi^{ij}\right)+2(K^{ik}K^j_k-\lambda K K^{ij})\nb\\
-\frac{1}{2}g^{ij}{\cal{L}}_K+\frac{1}{N}\nabla_k (\pi^{ik}N^j+\pi^{kj}N^i-\pi^{ij}N^k)\nb\\
-F^{ij}-F^{ij}_S-\frac{1}{2}g^{ij}{\cal{L}}_S-F^{ij}_a-\frac{1}{2}g^{ij}{\cal{L}}_A-F^{ij}_\varphi\nb\\
-\frac{1}{N}(AR^{ij}+g^{ij}\nabla^2A-\nabla^j\nabla^iA) =8\pi G
\tau^{ij},\;\;\;\;\;\;
 \eqn
where
 \bqn
 \lb{tauij}
\tau^{ij}&=&\frac{2}{\sqrt{g}N} \frac{\delta(\sqrt{g}N{\cal{L}}_M)}{\delta g_{ij}}, \nb\\
F_S^{ij}&=&-\sigma \left(\sigma_1a^ia^j+\sigma_2a^{ij}\right)\nb\\
                    &&+\frac{\sigma_1a^ka_k+\sigma_2a^k_k}{2}\left[(\nabla^i\varphi)(\nabla^j\varphi)+2\frac{N^{(i}\nabla^{j)}\varphi}{N}\right] \nb\\
                    &&+\frac{\sigma_2}{N}\nabla^{(i}[a^{j)}(A-{\cal
                    A})]-g^{ij}\frac{\sigma_2}{2N}\nabla^{k}[a_k(A-{\cal
                    A})],\nb\\
F^{ij}&=&\frac{1}{\sqrt{g}N}\frac{\delta (-\sqrt{g}N {\cal{L}}_V^R)}{\delta g_{ij}}
            =  \sum_{s=0}\hat{\gamma}_s\zeta^{n_s}(F_s)^{ij},\nb\\
F^{ij}_a&=&\frac{1}{\sqrt{g}N}\frac{\delta (-\sqrt{g}N {\cal{L}}_V^a)}{\delta g_{ij}}
               = \sum_{s=0}\hat\beta_s\zeta^{m_s}(F_s^a)^{ij},\nb\\
F^{ij}_\varphi&=&\frac{1}{\sqrt{g}N}\frac{\delta (-\sqrt{g}N {\cal{L}}_V^\varphi)}{\delta g_{ij}}
                          = \sum_{s=0}\mu_s(F_s^\varphi)^{ij},
 \eqn
 with $\hat\beta_s = (-\beta_0, \beta_n)\; (n = 1, 2, ..., 8)$, and
  \bqn
\hat{\gamma}_s &=& \left(\gamma_0, \gamma_1, \gamma_2, \gamma_3, \frac{1}{2}\gamma_5, -\frac{5}{2}\gamma_5,
3\gamma_5, \frac{3}{8}\gamma_5, \gamma_5, \gamma_5\right), \nb\\
n_s &=& (2, 0, -2, -2, -4, -4, -4, -4, -4, -4),\nb\\
m_s&=& (0, -2,-2,-2, -2, -2, -2, -2, -4 ), \nb\\
\mu_s &=& \left(2, 1, 1, 2, \frac{4}{3}, \frac{5}{3}, \frac{2}{3},
1-\lambda, 2-2 \lambda\right).
 \eqn

In addition, the matter components $(J^t, J^i, J_\varphi, J_A,
\tau^{ij})$ satisfy the conservation laws of energy and momentum,
\bqn
\label{energy conservation}
\int d^3x \sqrt{g} N \bigg[\dot{g}_{ij}\tau^{ij}-\frac{1}{\sqrt{g}}\partial_t (\sqrt{g} J^t)+\frac{2 N_i}{\sqrt{g} N}\partial_t (\sqrt{g} J^i)\nb\\
-\frac{A}{\sqrt{g} N}\partial_t (\sqrt{g} J_A)-2\dot{\varphi}J_\varphi\bigg]=0,\;\;\;\;\;\;\;\;\\
\label{mom conservation}
\frac{1}{N}\nabla^i(N\tau_{ik})-\frac{1}{\sqrt{g} N} \partial_t (\sqrt{g} J_k)-\frac{J_A}{2N}\nabla_k A-\frac{J^t}{2N}\nabla_kN\nb\\
-\frac{N_k}{N} \nabla_i J^i-\frac{J^i}{N}(\nabla_i N_k-\nabla_k
N_i)+J_\varphi \nabla_k\varphi=0.\;\;\;\;\;\;\;\; \eqn

As mentioned above, in writing down the general action (\ref{action}), we had added the ${\cal{L}}_{S}$ term, so that the action is the most general one with the enlarged symmetry
(\ref{symmetry}) and  nonprojectable condition $N = N(t, x)$. In addition, we had also flipped the sign of $g_{ij}$ in the $2\times 2$ matrix (\ref{gdbc}), so that the coupling constant $\beta_8$ becomes
non-negative now, in contrast to that given in \cite{ZWWS}\footnote{It should be noted that   the conclusions obtained in   \cite{ZWWS},    such as  stability, ghost,
and strong coupling, do not depend on the signs of   $\beta_8$, so they hold even after flipping the signs of $\beta_8$. This is also truce when applied the model to  the solar system
tests \cite{LW} and cosmology \cite{InflationB}.}.
In the following, we shall study the differences caused by these changes. In particular, we shall   consider their effects
on the existence of the spin-0 gravitons and  stability. To  anticipate, due to the presence of the term ${\cal{L}}_{S}$, spin-0 gravitons appear, although they are stable in a large
region of the phase space of the coupling constants. In addition, thanks to the  term $\beta_8 \left(\Delta{a^{i}}\right)^{2}$ in the potential (\ref{2.5a}), the theory is also
power-counting  renormalizable, even with the detailed balance condition. This is in contrast to all the versions with spin-0 gravitons  proposed so far.

\section {Spin-0 Gravitons and  Stability}
\renewcommand{\theequation}{3.\arabic{equation}} \setcounter{equation}{0}

When $\sigma_{1} = \sigma_2 = 0$, it was shown that the spin-0 gravitons are eliminated \cite{ZWWS}.
In this section, we consider the same problem for arbitrary $\sigma_{1}$ and $\sigma_{2}$.
Let us first note that the Minkowiski space-time,
$$
(\bar{N}, \bar{N}_{i}, \bar{g}_{ij}, \bar{A}, \bar{\varphi}) = (1, {0}, \delta_{ij}, 0, 0),
$$
 is a solution of the model with  $ \Lambda_g = \Lambda =0$,
  where quantities with bars denote the   background. Then,   the linear  scalar perturbations in the Minkowski background
 can be written in the form,
\bqn
N=1+\phi,\;\;N_i=\partial_i B,\nb\\
g_{ij}=(1-2 \psi) \delta_{ij}+2 \partial_i\partial_j E,\nb\\
A=\delta A,\;\;\;\;\varphi= \delta \varphi,
\eqn
where $\phi, B, \psi, E, \delta{A}$ and $\delta\varphi$ represent the scalar perturbations.
Using the gauge freedom, without loss of generality, we can always  choose the gauge \cite{ZWWS}
\bqn
E = 0 = \delta\varphi.
\eqn

To show our above claims given at the end of the last section, let us consider the quadratic action. After simple but tedious calculations,  we find that it takes the form,
\bqn
S^{(2)}&=&\zeta^2 \int dtd^3 x \Bigg\{(1-3 \lambda) (3 \dot{\psi}^2+2 \dot{\psi} \partial^2 B)\nb\\
&&+(1-\lambda) (\partial^2 B)^2-\left(\phi \eth + \frac{4 \beta_7 }{\zeta^2} \partial^2 \psi-\sigma_2 A\right)\partial^2\phi \nb\\
&& - \Big[2A-\gamma_1 (\psi-2\phi)+\alpha_1 \phi
\partial^2\Big]\partial^2 \psi\Bigg\}, ~~~~~~ \eqn where $\partial^2
= \delta^{ij}\partial_i \partial_j$, and
 \bqn
  \lb{gl.a}
\alpha_1 &\equiv& \frac{8 \gamma_2+3 \gamma_3}{\zeta^2},\nb\\
\lb{gl.b} \eth &\equiv& \beta_0 +\frac{\beta_2+\beta_4}{\zeta^2}
\partial^2- \frac{\beta_8}{\zeta^4} \partial^4.
\eqn

Now,  variations of $S^{(2)}$ with respect to $A$, $B$, $\phi$, and $\psi$ yield, respectively,
\bqn
\lb{eq.a1}
&&\partial^2\psi = \frac{1}{4}\sigma_2 \partial^2 \phi,\\
\lb{eq.a2}
&& (1-3 \lambda) \dot{\psi}+(1-\lambda)  \partial^2 B = 0,\\
\lb{eq.a3}
&& \eth \phi -\frac{\sigma_2}{2} A = 2 \wp \psi,\\
\lb{eq.a4}
&& \ddot{\psi}+\frac{1}{3} \partial^2 \dot{B} +\frac{2}{3(1-3 \lambda)} \Bigg\{(\alpha_1\partial^2-\gamma_1)\partial^2 \psi\nb\\
&&\;\;\;\;\;\;\;\;\;\;\;\;\;+\partial^2A+\wp \partial^2 \phi
\Bigg\}=0, \eqn where $\wp \equiv  -\gamma_1-\frac{\beta_7}{\zeta^2}
\partial^2$.

When $\sigma_{1} = \sigma_2 = 0$,  from Eq.(\ref{eq.a1}) it can be seen that  the scalar $\psi$ satisfies the Laplacian equation $\partial^2 \psi= 0$. Thus,
it does not represent a propagative mode, and with proper boundary conditions, one can always set it to zero. Similarly, this is also true for other scalars.
Hence,   the spin-0 gravitons are eliminated in this case, as shown in detail in \cite{ZWWS}.

When $\sigma_{1} \not= 0, \; \sigma_2 = 0$,  as one can see from Eqs.(\ref{eq.a1})-(\ref{eq.a4}), the above conclusion still holds.

However, when   $\sigma_{2} \not= 0$, the situation is different. In fact, from the above equations, we can obtain  a master equation for  the scalar mode $\psi$,
which in momentum space can be written in the form
\bqn
\lb{psiA}
\ddot{\psi}_k+\omega_k^2 \psi_k=0,
\eqn
where
\bqn
\lb{omegak}
\omega_k^2&=&\left(\frac{1-\lambda}{1-3\lambda}\right) \Bigg\{\left[\gamma_1\left(1-\frac{8}{\sigma_2}\right)-\frac{8 \beta_0}{\sigma_2^2}\right] k^2\nb\\
&&+\left[\alpha_1+\frac{8(\beta_2+\beta_4)}{\sigma_2 \zeta^2} + \frac{8\beta_7}{\sigma_2\zeta^2}\right]k^4+\frac{8 \beta_8}{\sigma_2^2 \zeta^4} k^6\Bigg\}.\nb\\
\eqn In the IR, the $k^2$ term dominates,  thus the stability of the
scalar mode in the IR requires
 \bqn
 \lb{a}
\left(\frac{1-\lambda}{1-3\lambda}\right)
\left[\gamma_1\left(1- \frac{8}{\sigma_2}\right)-\frac{8
\beta_0}{\sigma_2^2}\right]>0. \eqn In the UV,   the sixth order
derivative term dominates, and the stability now
requires \footnote{In the intermediate regime, by properly choosing
$\alpha_1, \beta_{2,4,7}$ and $\sigma_2$, one can always make
$\omega_k^2$ non-negative. This requirement yields very weak
constraints on these parameters. So, we do not need to consider them
explicitly here, but simply assume that  $\omega_k^2$ is
non-negative in this regime.}
 \bqn
 \lb{b}
 \left(\frac{1-\lambda}{1-3\lambda}\right) \beta_8 >0.
 \eqn

To study  the ghost problem, we consider the quadratic action $S^{(2)}$, which, after integrating out $\phi, B, A$, can be cast in the form,
 \bqn
S^{(2)}&=&\zeta^2 \int dtd^3x \Bigg\{2 \left(\frac{1-3\lambda}{1-\lambda}\right) \dot{\psi}^2\nb\\
&&\;\;\;\;\;- \left[\frac{16\beta_0}{\sigma_2^2}-2\gamma_1 \left(1-\frac{8}{\sigma_2}\right)\right] \psi\partial^2 \psi\nb\\
&&\;\;\;\;-\left[\frac{16 \beta_7}{\sigma_2 \zeta^2 }+\frac{16 (\beta_2+\beta_4)}{\sigma_2^2 \zeta^4}+2 \alpha_1\right]\psi \partial^2\psi\nb\\
&&\;\;\;\;\;+\frac{16 \beta_8}{\sigma_2^2 \zeta^4} \psi \partial^6 \psi\Bigg\}.
 \eqn
From the above one can see that the ghost-free condition reads
 \bqn
 \frac{1-3 \lambda}{1-\lambda}>0.
 \eqn
Then,  combining the stability conditions (\ref{a}) and (\ref{b}) with  this  ghost-free condition, we obtain
 \bqn
 \lb{conditionA}
 \beta_8 >0,\\
 \lb{conditionB}
i)\; \lambda > 1 \;\;\mbox{or}\;\; ii)\; \lambda< \frac{1}{3} ,\\
 \lb{conditionC}
 \sigma^{-}_{2} < \sigma_2 < \sigma^{+}_{2},
 \eqn
where
 \bq
 \lb{AA}
 \sigma_{2}^{\pm} \equiv 4\left\{-\gamma_1 \pm \sqrt{{\gamma}_{1}^2 - \frac{\beta_0}{2}}\right\}.
 \eq
Since $\sigma_2$ is real, the condition (\ref{conditionC}) holds only when
 \bq
 \lb{AB}
 {\gamma}_{1}^2 - \frac{\beta_0}{2} \ge 0.
 \eq

\section {Coupling with Matter}
\renewcommand{\theequation}{4.\arabic{equation}} \setcounter{equation}{0}

Before studying the post-Newtonian limits, we must first specify the coupling between matter and the HL gravity.
This has not been done systematically in the HL theory \cite{reviews}.

As shown in \cite{Lin:2012bs}, the projectable version of the $U(1)$
extension of HL gravity can be consistent with the
solar system tests only if the gauge field $A$ and
the Newtonian prepotential $\varphi$ couple to matter fields in such a
way that
\begin{equation}
 {\cal N} \equiv N \left[1-\upsilon\sigma+O(\sigma^2)\right],
  \label{eqn:geometrical-lapase}
\end{equation}
(with $\upsilon\simeq 1$) plays the role of the geometrical lapse function,
where $\sigma$ is defined by (\ref{2.6}). In the present paper we shall
see that this kind of prescription for the coupling of the Newtonian
prepotential to matter fields can be also generalized to
the non-projectable version \footnote{In \cite{LW} the static spherical vacuum spacetimes were
studied in the IR in the nonprojectable case with the enlarged symmetry (\ref{symmetry}), and
found that  the theory is consistent with the solar system tests even without $A$ and $\varphi$
being part of the metric.}. As already pointed out in \cite{Lin:2012bs},
from UV viewpoints, it is not obvious how to obtain such a prescription
from the action principle. Actually, in the UV, $\sigma$ has a
non-vanishing scaling dimension and thus it is not easy to imagine how a
linear combination of $N$ and $N\sigma$ can universally enter the UV
action of matter fields. On the other hand, in the IR, $\sigma$ is
dimensionless and thus, this kind of prescription is not forbidden a
priori. In Appendix C, we consider a scalar-tensor extension of the
$U(1)$ extension of HL  gravity (with or without the
projectability condition) to make it possible for the prescription
(\ref{eqn:geometrical-lapase}) to emerge in the IR, with the expense of
fine-tuning in the IR but without spoiling the power-counting
renormalizability of the theory in the UV.

In the scalar-tensor extension elaborated in Appendix C, it is possible
that in the IR, matter fields universally couple to the ADM components
$(\tilde{N}, \tilde{N}^i, \tilde{g}_{ij})$ defined as
\bqn
\lb{eq8-1}
& & \tilde{N} = F(\sigma)N,\;\;\;
\tilde{N}^i = N^i + Ng^{ij} \nabla_j\varphi,\nb\\
&& 
\tilde{g}_{ij} = \Omega^2(\sigma)g_{ij},
\eqn
with
\bqn
\lb{eq8-1a}
& & F = 1 - a_1\sigma, \;\;\;
 \Omega = 1 - a_2\sigma,
\eqn
where $a_1$ and $a_2$ are two arbitrary coupling constants. Note that by
setting the first terms in $F$ and $\Omega$ to unity, we have used the
freedom to rescale the units of time and space. Therefore, the parameter
$\gamma_1$ for example can no longer be rescaled and thus its value has
a physical meaning by itself. For later convenience we also define
\bq
\lb{eq8-2}
\tilde{N}_i =\Omega^2(\sigma)\left(N_i + N\nabla_i\varphi\right),\;\;\;
\tilde{g}^{ij} = \Omega^{-2}(\sigma)g^{ij}.
\eq

Since matter fields universally couple to ($\tilde{N}$, $\tilde{N}^i$,
$\tilde{g}_{ij}$), the matter action is of the form
\bqn
\lb{eu7}
S_{m} &=& \int{dtd^3x \tilde{N}\sqrt{\tilde{g}}\;  \tilde{\cal{L}}_{m}\left(\tilde{N}, \tilde{N}_i, \tilde{g}_{ij}; \psi_{n}\right)},
\eqn
where $\psi_n$ collectively  stands for matter fields. One can then define the matter
stress-energy in the ADM decomposition as
\bqn
\lb{eq11}
 \rho_{H}  &\equiv& - \frac{\delta(\tilde{N}\tilde{\cal{L}}_{m})}{\delta\tilde{N}}, \nb\\
s^i &\equiv&  \frac{\delta(\tilde{N}\tilde{\cal{L}}_{m})}{\delta\tilde{N}_i}, \nb\\
s^{ij} &\equiv& \frac{2}{\tilde{N} \sqrt{\tilde{g}}}   \frac{\delta(\tilde{N} \sqrt{\tilde{g}}\tilde{\cal{L}}_{m})}{\delta\tilde{g}_{ij}},
\eqn
so that
\bqn
\lb{eq10}
\delta_{\tilde{N}}S_{m} & = &
- \int{dt d^3x \sqrt{\tilde{g}} \rho {\delta\tilde{N}}},\nb\\
\delta_{\tilde{N}_i}S_{m} &=&
\int{dt d^3x \sqrt{\tilde{g}} s^i{\delta\tilde{N}_i}},\nb\\
\delta_{\tilde{g}_{ij}}S_{m} &=&
\frac{1}{2} \int{dt d^3x \tilde{N} \sqrt{\tilde{g}} s^{ij}{\delta\tilde{g}_{ij}}}.
\eqn
On the other hand, the source terms in gravity equations of motion
are defined through variation of the matter action w.r.t. $N$, $N_i$,
$g_{ij}$, $A$ and $\varphi$. In the following, we express those source
terms in terms of the components of the matter stress energy shown in
(\ref{eq11}).

From Eqs.(\ref{eq8-1}) and (\ref{eq8-2}), we find that
\bqn
\lb{eq12}
\tilde{N} &=&\tilde{N}(N, N_i, g_{ij}, A, \varphi),\nb\\
\tilde{N}_i &=&\tilde{N}_i(N, N_i, g_{ij}, A, \varphi),\nb\\
\tilde{g}_{ij} &=&\tilde{g}_{ij}(N, N_i, g_{ij}, A, \varphi).
\eqn
Therefore,
\bqn
\lb{eq13}
\delta_{{N}}S_{m} &=&\int{dtd^3x\Bigg\{\frac{\delta\tilde{N}}{\delta{N}}\frac{\delta(\tilde{N}\sqrt{\tilde{g}}\tilde{\cal{L}}_{m})}{\delta\tilde{N}}
+ \frac{\delta\tilde{N}_i}{\delta{N}}\frac{\delta(\tilde{N}\sqrt{\tilde{g}}\tilde{\cal{L}}_{m})}{\delta\tilde{N}_i}}\nb\\
&& + \frac{\delta\tilde{g}_{ij}}{\delta{N}}\frac{\delta(\tilde{N}\sqrt{\tilde{g}}\tilde{\cal{L}}_{m})}{\delta\tilde{g}_{ij}}\Bigg\}\delta{N}\nb\\
&=& \int{dtd^3x\sqrt{g}\Omega^3(\sigma)\Bigg\{- \rho_H
\frac{\delta\tilde{N}}{\delta{N}}
+  \frac{\delta\tilde{N}_i}{\delta{N}}s^i }\nb\\
&&  + \frac{1}{2}\tilde{N}  \frac{\delta\tilde{g}_{ij}}{\delta{N}}s^{ij}\Bigg\}\delta{N}} \equiv \frac{1}{2}\int{dt d^3x \sqrt{g} J^t \delta{N},
\eqn
or
\bqn
\lb{eq14}
J^{t} &=& 2\Omega^{3}(\sigma)\Bigg\{- \rho_H  \frac{\delta\tilde{N}}{\delta{N}}
+  \frac{\delta\tilde{N}_i}{\delta{N}}s^i   + \frac{1}{2}\tilde{N}  \frac{\delta\tilde{g}_{ij}}{\delta{N}}s^{ij}\Bigg\}. ~~~~~~~~~
\eqn
Similarly, it can be shown that
\bqn
\lb{eq15}
J^{i} &=& - \Omega^{3}(\sigma)\Bigg\{- \rho_H  \frac{\delta\tilde{N}}{\delta{N}_i}
+  \frac{\delta\tilde{N}_k}{\delta{N}_i}s^k   + \frac{1}{2}\tilde{N}  \frac{\delta\tilde{g}_{kl}}{\delta{N}_i}s^{kl}\Bigg\},\nb\\
\tau^{ij} &=& \frac{2 \Omega^{3}(\sigma)}{N}\Bigg\{- \rho_H
\frac{\delta\tilde{N}}{\delta{g}_{ij}}
+  \frac{\delta\tilde{N}_k}{\delta{g}_{ij}}s^k   + \frac{1}{2}\tilde{N}  \frac{\delta\tilde{g}_{kl}}{\delta{g}_{ij}}s^{kl}\Bigg\},\nb\\
J_{A} &=& 2 \Omega^{3}(\sigma)\Bigg\{- \rho_H
\frac{\delta\tilde{N}}{\delta{A}}
+  \frac{\delta\tilde{N}_k}{\delta{A}}s^k   + \frac{1}{2}\tilde{N}  \frac{\delta\tilde{g}_{kl}}{\delta{A}}s^{kl}\Bigg\},\nb\\
J_{\varphi} &=& - \frac{1}{N}\Bigg\{\frac{1}{\sqrt{g}}\left(B \sqrt{g}\right)_{,t} - \nabla_{i}\Big[B\left(N^i + N \nabla^i\varphi\right)\Big]\nb\\
&& ~~~~~~~~~ - \nabla_i\left(N\Omega^5 s^i\right)\Bigg\},
\eqn
where
\bqn
\lb{eq16}
B &\equiv& - \Omega^{3}(\sigma)\Bigg\{a_1\rho_H  - \frac{2a_2\left(1- a_2\sigma\right)}{N}s^k\left(N_k + N\nabla_k\varphi\right)\nb\\
&& -  a_2\left(1- a_1\sigma\right)\left(1- a_2\sigma\right)g_{ij}s^{ij}\Bigg\}.
\eqn

On the other hand, from Eqs.(\ref{eq8-1}) and (\ref{eq8-2}), we find
that
\bqn
\lb{eq17}
 \frac{\delta\tilde{N}}{\delta{N}} &=& 1 + \frac{1}{2}a_1 \left(\nabla_k\varphi\right)^2, \nb\\
 \frac{\delta\tilde{N}_i}{\delta{N}} &=& \frac{\Omega}{N}\Bigg\{N\Omega\nabla_{i}\varphi \nb\\
 && ~~~~~~ + 2a_2\left(N_i + N\nabla_i\varphi\right)
 \left[\sigma + \frac{1}{2}\left(\nabla_k\varphi\right)^2\right]\Bigg\},\nb\\
  \frac{\delta\tilde{g}_{ij}}{\delta{N}} &=& \frac{2a_2\Omega}{N} \left[\sigma + \frac{1}{2}\left(\nabla_k\varphi\right)^2\right]g_{ij},\nb\\
   \frac{\delta\tilde{N}}{\delta{N}_i}  &=& a_1\nabla^i\varphi,\nb\\
   \frac{\delta\tilde{N}_k}{\delta{N}_i}  &=& \frac{\Omega}{N}\Big\{N\Omega\delta^{i}_{k}
   + 2a_2\left(N_k + N\nabla_k\varphi\right)\nabla^i\varphi\Big\},\nb\\
\frac{\delta\tilde{g}_{kl}}{\delta{N}_i}    &=& \frac{2a_2\Omega}{N}g_{kl}\nabla^i\varphi,\nb\\
\frac{\delta\tilde{N}}{\delta{g}_{ij}} &=& -a_1\Bigg[N^{(i}\nabla^{j)}\varphi + \frac{1}{2}N \left(\nabla^i\varphi\right)\left(\nabla^j\varphi\right)\Bigg],\nb\\
\frac{\delta\tilde{N}_k}{\delta{g}_{ij}} &=&-\frac{2a_2\Omega}{N}\left(N_k + N\nabla_k\varphi\right)\nb\\
&& \times \Bigg[N^{(i}\nabla^{j)}\varphi + \frac{1}{2}N \left(\nabla^i\varphi\right)\left(\nabla^j\varphi\right)\Bigg],\nb\\
 \frac{\delta\tilde{g}_{kl}}{\delta{g}_{ij}} &=& \frac{1}{2}\Omega^2\left(\delta^i_k\delta^j_l  + \delta^i_l\delta^j_k\right)\nb\\
 && - \frac{2a_2\Omega g_{kl}}{N} \Bigg[N^{(i}\nabla^{j)}\varphi + \frac{1}{2}N \left(\nabla^i\varphi\right)\left(\nabla^j\varphi\right)\Bigg],\nb\\
  \frac{\delta\tilde{N}}{\delta{A}} &=& -a_1,\nb\\
  \frac{\delta\tilde{N}_i}{\delta{A}} &=& - \frac{2a_2\Omega}{N}\left(N_i + N\nabla_i\varphi\right),\nb\\
    \frac{\delta\tilde{g}_{ij}}{\delta{A}} &=& - \frac{2a_2\Omega}{N}g_{ij}.
\eqn

For the gauge $\varphi = 0$, the above expressions reduce to
\bqn
\lb{eq18}
 \frac{\delta\tilde{N}}{\delta{N}} &=& 1, \;\;\;\;
 \frac{\delta\tilde{N}_i}{\delta{N}} = \frac{2a_2\sigma \Omega}{N} N_i,\;\;\;
  \frac{\delta\tilde{g}_{ij}}{\delta{N}} = \frac{2a_2\sigma \Omega}{N}  g_{ij},\nb\\
   \frac{\delta\tilde{N}}{\delta{N}_i}  &=& 0,\;\;\;
   \frac{\delta\tilde{N}_k}{\delta{N}_i}  =  \Omega^2\delta^{i}_{k},\;\;\;
\frac{\delta\tilde{g}_{kl}}{\delta{N}_i}    = 0,\;\;\;
\frac{\delta\tilde{N}}{\delta{g}_{ij}} = 0,\nb\\
\frac{\delta\tilde{N}_k}{\delta{g}_{ij}} &=&0,\;\;\;
 \frac{\delta\tilde{g}_{kl}}{\delta{g}_{ij}} = \frac{1}{2}\Omega^2\left(\delta^i_k\delta^j_l  + \delta^i_l\delta^j_k\right),\nb\\
  \frac{\delta\tilde{N}}{\delta{A}} &=& -a_1,\;\;\;
  \frac{\delta\tilde{N}_i}{\delta{A}} = - \frac{2a_2\Omega}{N}N_i,\nb\\
    \frac{\delta\tilde{g}_{ij}}{\delta{A}} &=& - \frac{2a_2\Omega}{N}g_{ij},\; \; (\varphi = 0).
\eqn
Inserting the above expressions into Eq.(\ref{eq15}), we find that
\bqn
\lb{eq19}
J^{t} &=& 2\Omega^{3}\Bigg\{- \rho_H
+  \frac{2a_2\sigma\Omega}{N}N_is^i   + a_2\sigma\Omega(1-a_1\sigma)g_{ij}s^{ij}\Bigg\},\nb\\
J^{i} &=& - \Omega^{5}s^i,\nb\\
\tau^{ij} &=& (1-a_1\sigma)\Omega^5s^{ij},\nb\\
J_{A} &=& 2 \Omega^{3}\Bigg\{a_1 \rho_H  -\frac{2a_2\Omega}{N}N_ks^k  -a_2\Omega(1-a_1\sigma)g_{ij}s^{ij}\Bigg\},\nb\\
J_{\varphi} &=& - \frac{1}{N}\Bigg\{\frac{1}{\sqrt{g}}\left(B \sqrt{g}\right)_{,t} - \nabla_{i}\Big[B\left(N^i + N \nabla^i\varphi\right)\Big]\nb\\
&& ~~~~~~~~~ - \nabla_i\left(N\Omega^5 s^i\right)\Bigg\},\; (\varphi = 0).
\eqn

 \section {Post-Newtonian Approximations}
\renewcommand{\theequation}{5.\arabic{equation}} \setcounter{equation}{0}

In the low energy limit, the high-order derivative terms are highly suppressed, and can be safely dropped out, so the action reduces to
 \bqn
 \label{4.1}
S_{IR}&=&\zeta^2 \int dtd^3x \sqrt{g} N \Big({\cal{L}}_{K}-{{\cal{L}}}^{IR}_V + {{\cal{L}}}_{A} \nb\\
&& ~~~~~~~~~~~~~~~~+ {\cal{L}}_{S}+ {\cal{L}}_{\varphi}+
{\zeta^{-2}} {\cal{L}}_M\Big),
 \eqn
where
 \bqn
 \lb{4.2}
 {{\cal{L}}}_V^{IR}&=& 2\Lambda + \gamma_1 R-\beta_0a_ia^i.
 \eqn
Note that, for the action  to have the U(1) symmetry even in the IR, here we have kept all the terms in both ${\cal{L}}_{\varphi}$ and $ {{\cal{L}}}_{A}$.
Then, using this U(1) gauge freedom, we set
\bq
\lb{4.3}
 \varphi=0,
 \eq
which  uniquely fix the gauge \cite{ZWWS}. On the other hand, in the solar system the influence of the cosmological
 constant and the space curvature are negligible. So, in this section we can safely set
\bq
\lb{4.4}
\Lambda = \Lambda_g = 0.
\eq

With the above gauge choice, it can be shown that   the Hamiltonian,  momentum constraints,  and the dynamical  equations,
can be cast, respectively, in the forms,
 \bqn
 \lb{4.5a}
&& K_{ij}K^{ij}-\lambda K^2 + \gamma_1 R +\beta_0(2a^i_{\;\; i}+a_i a^i) \nb\\
&& ~~~~~~~~~~~~+2\frac{\sigma_1}{N}\nabla_i\left(Aa^i\right)-\frac{\sigma_2}{N}\nabla^2A = 8\pi G J^t,  \\
 \lb{4.5b}
&& \nabla^k{\pi}_{ik}= 8\pi G J_i,\\
 \lb{4.5c}
&& \frac{g_{il}g_{jk}}{N\sqrt{g}}\frac{\partial}{\partial t}\left(\sqrt{g}\; {\pi}^{lk}\right)+2(K_{ik}K^k_j-\lambda KK_{ij})\nb\\
&& ~~~~~~~  +\frac{1}{N}\nabla_k\left({\pi}^k_iN_j+{\pi}^k_jN_i-{\pi}_{ij}N^k\right)\nb\\
&& ~~~~~~~  -\frac{1}{2}g_{ij}\left(K_{lk}K^{lk}-\lambda K^2\right)-{\cal F}_{Sij}-\gamma_1{\cal{F}}_{ij}\nb\\
&&~~~~~~-\frac{A}{2N}g^{ij}(\sigma_1a^ka_k+\sigma_2a^k_k)-{\varepsilon}_{ij}
=8\pi G \tau_{ij},\;\;\;\;\;\;
 \eqn

 where
  \bqn \lb{4.6}
{\varepsilon}_{ij}&\equiv& -\frac{{A}R}{2N}g_{ij} -\beta_0\left(a_ia_j-\frac{1}{2}a_ka^k g_{ij}\right)\nb\\
&&+\frac{1}{N}\left(AR_{ij}+g_{ij}\nabla^2A -\nabla_j\nabla_iA\right),\nb\\
{\cal{F}}_{ij}&\equiv&R_{ij}-\frac{1}{2}g_{ij}R+\frac{1}{N}\left(g_{ij}\nabla^2N-\nabla_j\nabla_iN\right),\nb\\
{\cal{F}}_{Sij}&\equiv&-\frac{A}{N} \left(\sigma_1a_ia_j+\sigma_2a_{ij}\right)\nb\\
                    &&+\frac{\sigma_2}{N}\nabla_{(i}(a_{j)}A)-g_{ij}\frac{\sigma_2}{2N}\nabla^{k}(a_kA)\nb\\
{\pi}_{ij}&\equiv&-K_{ij}+\lambda Kg_{ij}.
 \eqn

Note that   the energy-momentum tensor   in GR is defined as,
\bq
\lb{4.7}
T^{\mu\nu} = \frac{1}{\sqrt{-\gamma}} \frac{\delta\left(\sqrt{-\gamma}{\cal{L}}_{M}\right)}{\delta \gamma_{\mu\nu}},
\eq
where $\gamma_{\mu\nu}$ is given by,
\bqn
\lb{Pmetric}
\left(\gamma_{\mu\nu}\right) &=& \left(\matrix{-\tilde{N}^2 + \tilde{N}^{i}\tilde{N}_{i} & \tilde{N}_{i}\cr
\tilde{N}_{i} & \tilde{g}_{ij}\cr}\right),\nb\\
\left(\gamma^{\mu\nu}\right) &=&
\left(\matrix{-\frac{1}{\tilde{N}^2}   &
\frac{\tilde{N}^{i}}{\tilde{N}^2}\cr
 \frac{\tilde{N}^{i}}{\tilde{N}^2} & \tilde{g}^{ij} - \frac{\tilde{N}^{i}\tilde{N}^{j}}{\tilde{N}^2}\cr}\right),
\eqn
with $\tilde{g}^{ij}\tilde{g}_{ik} = \delta^{j}_{k}$ and $\tilde{N}_{i} \equiv \tilde{g}_{ij}\tilde{N}^{j}$.

Introducing the normal vector $n_{\nu}$ to the hypersurface $ t =$ Constant,
\bq
\lb{4.8}
n_{\mu} = -\tilde{N}\delta^{t}_{\mu}, \;\;\; n^{\mu} =
\frac{1}{\tilde{N}} \left(1,  -\tilde{N}^{i} \right),
 \eq
one can decompose $T_{\mu\nu}$ as \cite{PA01},
\bqn
\lb{4.9}
\rho_H  &\equiv& T_{\mu\nu} n^{\mu} n^{\nu},\nb\\
s_{i}  &\equiv&  -  T_{\mu\nu} \left(\delta^{\mu}_{i} + n^{\mu}n_{i}\right)   n^{\nu},\nb\\
s_{ij}  &\equiv&  T_{\mu\nu}  \left(\delta^{\mu}_{i} +
n^{\mu}n_{i}\right)   \left(\delta^{\nu}_{j} +
n^{\nu}n_{j}\right),
 \eqn
  in terms of which, the quantities
$J^t, \; J_i, \; J_A, \; J_{\varphi}$ and $\tau_{ij}$ are given by
Eq.(\ref{eq15}).

On the other hand,  the variations of $S_{IR}$ with respect to $\varphi$ and $A$ yield,
\bqn
\lb{4.17a}
&& {\cal G}^{ij}K_{ij}-\frac{\hat{\cal G}^{ijlk}}{N}\Big[\nabla_l\left(a_k N K_{ij}\right)  +\nabla_k\left(a_l N K_{ij}\right)\Big]\nb\\
&&-\frac{1}{2N}\left[\frac{1}{\sqrt{g}}\frac{\partial}{\partial
t}\left(\sqrt{g}Z_A\right)-\nabla_j\left(N^jZ_A\right)\right]\nb\\
&& +\frac{{\tilde\lambda}}{N}\left[\nabla^2\left(NK\right)-\nabla^i\left(a_i N K\right)\right]= 8\pi G J_{\varphi},\\
\lb{4.17b} && R-Z_A = 8\pi G J_{A},
 \eqn
where
 \bqn \lb{4.17c}
Z_A=\sigma_1a_ia^i+\sigma_2a^i_i.
 \eqn
 Eqs.(\ref{4.5a})-(\ref{4.17c}) are the HL field equations in the IR limit.

For a   perfect fluid, a fundamental equation
is the conservation of the baryon number $n$ \cite{Brown},
\bq
\lb{4.18}
\frac{\partial{n}}{\partial t} + \vec{\nabla}\centerdot (n\vec{v}) = 0,
\eq
where $\vec{v}$ is the three-velocity of the fluid. Introducing the rest mass density $\rho$ of the atoms in the element of the fluid by
$\rho = \mu n$, where $\mu$ is the mean rest mass per baryon in the element, the above equation can be written in the form,
\bq
\lb{4.19}
 \left(\frac{\partial}{\partial t} + \vec{\nabla}\centerdot \vec{v}\right)\rho = 0.
\eq

 In the post-Newtonian approximations, we assume that the metric can be written in the form \cite{Will},
\bq
\lb{4.25}
\gamma_{\mu\nu} = \eta_{\mu\nu} + h_{\mu\nu},
\eq
where $\eta_{\mu\nu} = {\mbox{diag.}}\left(-1, 1, 1, 1\right)$, and
\bqn
\lb{4.26}
h_{00}&\sim&{\cal O}(2)+{\cal O}(4),\nb\\
h_{0j}&\sim&{\cal O}(3),\nb\\
h_{ij}&\sim&{\cal O}(2)+{\cal O}(4),
\eqn
where ${\cal O}(n) \equiv {\cal O}\left(v^n\right)$. It should be noted
that, in contrast to GR, here $h_{ij}$ needs to be expanded to the
fourth-order of $v$ in order  to obtain consistent field equations for
the Hamiltonian constraint, $A$-constraint, and the trace part of the
dynamical equations. Generalizing the arguments presented in
\cite{Will} to the present case, we find that, up to ${\cal{O}}(4)$
order, $h_{\mu\nu}$ in the post-Newtonian approximations consists of
the terms,
\bqn
\lb{4.20}
&& h_{00}\left[{\cal{O}}(4)\right]: \; U^2, \Phi_{W}, \Phi_{1},  \Phi_{2},  \Phi_{3},  \Phi_{4},  \mathfrak{A},  \mathfrak{B} \nb\\
&& h_{0i}\left[{\cal{O}}(3)\right]: \; V_{i}, W_{i}\nb\\
&& h_{ij}\left[{\cal{O}}(4)\right]: \; U\delta_{ij},  \chi_{,ij},\; h_{ij}^{(4)},
\eqn
where $U, \Phi_{W}, ..., \chi$ are given by Eq.(\ref{4.21}) in Appendix D, and $h_{ij}^{(4)} \simeq {\cal O}(4)$. From the continuity equation (\ref{4.19}), one can
obtain various useful relations, part of which is listed in Eq.(\ref{4.24}).

Under the gauge transformations (\ref{A.0e}), we find that
\bqn
\lb{4.22}
\bar{h}_{\bar{0}\bar{0}}&=&h_{00}-2\lambda_2(U^2+\Phi_W-\Phi_2)+2\dot{\xi_0}\nb\\
\bar{h}_{\bar{0}\bar{j}}&=&h_{0j}-\lambda_2\chi_{,0j}, \nb\\
\bar{h}_{\bar{i}\bar{j}}&=&h_{ij}-2\lambda_2\chi_{,ij},
\eqn
where in writing the above expressions, we had chosen    $\xi_j=\lambda_2\chi_{,j}$ \cite{Will} with $\lambda_2$ being
an arbitrary  constant. Clearly, by properly choosing $\lambda_2$ we can eliminate the anisotropic term $\chi_{,ij}$, as it was done in the {\em standard
post-Newtonian gauge} \cite{Will}. However, since now $\xi_0$ is a function of $t$ only, we cannot eliminate the  $\mathfrak{B}$ term in $h_{00}$.
Therefore, the general metric coefficients up to   ${\cal{O}}(4)$ order in the HL theory are given by,
\bqn
\lb{4.23}
\gamma_{00} &=& -1 + 2U - 2\beta U^2 - 2\xi \Phi_{W} \nb\\
&& + \left(2+ 2\gamma + \alpha_3 + \zeta_1 - 2\xi\right)\Phi_{1}\nb\\
&& + 2\left(1+3\gamma - 2\beta + \zeta_2 + \xi\right)\Phi_2 \nb\\
&& + 2\left(1+\zeta_3\right)\Phi_3 + 2\left(3\gamma + 3\zeta_4 -2\xi\right)\Phi_4\nb\\
&& -\left(\zeta_1 - 2\xi\right) \mathfrak{A} + \zeta_{B} \mathfrak{B},\nb\\
\gamma_{0i} &=& - \frac{1}{2} \left(3 + 4\gamma + \alpha_1 - \alpha_2 + \zeta_1 - 2\xi\right)V_{i} \nb\\
&& - \frac{1}{2}\left(1 + \alpha_2 - \zeta_1 + 2\xi\right) W_{i},\nb\\
\gamma_{ij} &=& \left(1 + 2\gamma U\right)\delta_{ij} + h^{(4)}_{ij},
\eqn
where $\beta, \gamma, \xi, \zeta_{i}, \alpha_{j}$ and $\zeta_{B}$ are the eleven independent constants that characterize the post-Newtonian limits in the HL theory 
\footnote{Note that in theories with general covariance, one can always use the gauge freedom of $\xi_0$, which is now a function of both $t$ and $x$,  to eliminate the term
$\mathfrak{B}$ in $\gamma_{00}$, so finally there are only ten independent constants \cite{Will}.}. Thus, $h_{\mu\nu}$ can be written in the form,
\bqn
\lb{hmunu}
h_{00}&=&2U+{\cal O}(4),\nb\\
h_{0j}&=& cV_j+d\chi_{,0j}+{\cal O}(5),\nb\\
h_{ij}&=& 2\gamma U\delta_{ij}+b\chi_{,ij}  +{\cal O}(4).
\eqn
Using the gauge freedom mentioned above, we shall set $b = 0$.
In the following, we shall solve the HL equations (\ref{4.5a})-(\ref{4.5c}) and (\ref{4.17a})-(\ref{4.17b}) order by order.

i) Hamiltonian constraint to ${\cal O}(2)$: To this order,  from
Eq.(\ref{eq19}) we find that $J^t = -2\rho$,  and Eq.(\ref{4.5a})
takes the form,
 \bqn
 \lb{4.28}
 &&\Big(2\beta_0a_1-4\gamma_1a_2
-\sigma_2\Big) \partial^2A_2\nb\\
&& ~~~~~~~~~~
 =2(2\gamma_1\gamma + \beta_0+ 2\varkappa)\partial^2U,  ~~~~~~~
 \eqn
where $\varkappa \equiv G/G_{N}$, and $G_{N}$ is the Newtonian constant defined in Eq.(\ref{4.24}).
Then, Eq.(\ref{4.28})  yields  $A_2 = f U$,  where
\bqn
\lb{4.30}
f \equiv \frac{2\beta_0+4\gamma_1\gamma+4\varkappa}{2\beta_0a_1-4\gamma_1a_2-\sigma_2}.
\eqn
Note that in writing the above expression, we had expanded $A$ as \cite{Will}
\bq
\lb{4.29}
A =    A_2 + A_4,
\eq
where $A_n$ is of order ${\cal{O}}(n)$.   This is consistent with the results obtained in the spherical case \cite{LW}.

ii) Momentum constraint to ${\cal O}(3)$: To this order, we find that  $J_i = -2\rho v_i$, and Eq.(\ref{4.5b}) becomes
 \bqn
 \lb{4.31}
&& \Big[\frac{c}{2}+2d-b+\gamma+a_2f-\lambda(c+3\gamma+2d-b)\nb\\
&& ~~ -3\lambda a_2f\Big]U_{,0i} -\frac{c}{2}\partial^2V_i
=2\varkappa\partial^2V_i.\nb
 \eqn
Therefore, we obtain $c = -4\varkappa$, and
\bqn
\lb{4.32}
&&  \gamma+ 2d+ (1 - 3\lambda) a_2f  -\lambda(3\gamma+2d - 4\varkappa) \nb\\
&& ~~~~~~~~~  - 2\varkappa = 0.
\eqn

iii) Dynamical equations to ${\cal O}(2)$: In this order, we find
$\tau_{ij} = 0$, and Eq.(\ref{4.5c}) becomes
 \bqn
 \lb{4.34a}
\Big[f-\gamma_1(\gamma+a_2f)+\gamma_1(1-a_1f)\Big]\left(U_{,ij}-\delta_{ij}\partial^2U\right)=0,\nb
 \eqn from which we obtain
  \bq
  \lb{4.35}
f+\gamma_1(\gamma+a_2f)-\gamma_1(1-a_1f)=0.
 \eq

iv)  $A$-constraint  to ${\cal O}(2)$: In this order, we find that  $J_A = 2a_1\rho$, and that Eq.(\ref{4.17b}) becomes
 \bqn
 \lb{4.36}
 \left[\sigma_2(1-a_1f)-4(\gamma+a_2f)\right]\partial^2U   = -4\varkappa a_1\partial^2U. \nb
 \eqn
Therefore, we have
 \bqn
 \lb{4.36a}
4\varkappa a_1-4(\gamma+a_2f)+\sigma_2(1-a_1f)=0.
 \eqn
v)  $\varphi$-constraint  to ${\cal O}(3)$: In this order, we find that  $J_\varphi = a_1\rho_{,0}+(\rho v_k)_{,k}$ and Eq.(\ref{4.17b}) becomes
 \bqn
 \lb{4.36ba}
&& \Big[(1-\lambda)(3\lambda+2d+3a_2f - 4\varkappa)-\frac{\sigma_2}{2}(1-a_1f)\Big]\partial^2U_{,0}\nb\\
&& ~~~~~~~~~~~~~~~~ = 2\varkappa (a_1-1)\partial^2U_{,0}, \nb
 \eqn
from which we obtain
\bqn
\lb{4.36b}
&& (1-\lambda)(3\lambda+2d+3a_2f - 4\varkappa)\nb\\
&&~~~~~~   - 2\varkappa (a_1-1)  -\frac{\sigma_2}{2}(1-a_1f) = 0.
\eqn
However, this equation is not independent. In fact, it can be obtained from   Eqs.(\ref{4.32}), (\ref{4.35}) and  (\ref{4.36a}).

vi)  The Hamiltonian constraint, $A$-constraint  and the trace of dynamical equations to ${\cal O}(4)$: To the fourth-order of $v$, these three equations contain
 $R\left(h^{(4)}_{ij}\right)$ and $\partial^2A_4$ terms. In particular, from the trace part of the dynamical equations and the $A$-constraint, we can eliminate $R$ to obtain an equation that contains $\partial^2A_4$. Combining such obtained equation with the Hamiltonian constraint, we can further eliminate the $\partial^2A_4$ term, so finally
a differential  equation involved only $h_{00}$ is obtained.  For more details, see Appendix E. 
Then, after solving the resulted  equation, we find that  $h_{00}$ is given by
\bqn
\lb{4.37}
h_{00}&=&2U + h_1({\mathfrak A}+{\mathfrak B})+ h_2\Phi_1 + h_3 \Phi_2 + h_4\Phi_3\nb\\
&&+h_5\Phi_4+ h_6 U^2,
\eqn
where $h_{n}$ are given by Eq.(\ref{4.37a}) in Appendix E.
Comparing the above expression with Eq.(\ref{4.23}), and considering the fact that $\gamma_{00} = -1 + h_{00}$, we find that
\bqn
\lb{4.38a}
&& -2\beta = h_6,\\
\lb{4.38b}
&& - 2\xi = 0,\\
\lb{4.38c}
&& 2+ 2\gamma + \alpha_3 + \zeta_1 - 2\xi = h_2,\\
\lb{4.38d}
&& 2\left(1+3\gamma - 2\beta + \zeta_2 + \xi\right) = h_3,\\
\lb{4.38e}
&& 2\left(1+\zeta_3\right) = h_4,\\
\lb{4.38f}
&& 2\left(3\gamma + 3\zeta_4 -2\xi\right) = h_5,\\
\lb{4.38g}
&&  -\left(\zeta_1 - 2\xi\right) = h_1,\\
\lb{4.38h}
&&  \zeta_{B} = h_1.
\eqn
 Combining Eqs.(\ref{4.32}), (\ref{4.35}), (\ref{4.36a}) and (\ref{4.38a})-(\ref{4.38h}), we finally obtain the eleven independent PPN
parameters, which are given by Eq.(\ref{4.39}) in Appendix F.

Current experiment limits on the PPN parameters are given by  \cite{Will,RJ,PF13},
\bqn
\lb{4.47}
\gamma&=&1+(2.1\pm2.3)\times10^{-5},\nb\\
\beta&=&1+(-4.1\pm7.8)\times10^{-5},\nb\\
\alpha_1&<&10^{-4},\nb\\
\alpha_2&<&4\times10^{-7},\nb\\
\alpha_3&<&4\times10^{-20},\nb\\
\xi&<&10^{-3},\nb\\
\Gamma&<&1.5\times10^{-3},\nb\\
\zeta_1&<&2\times10^{-2},\nb\\
\zeta_2&<&4\times10^{-5},\nb\\
\zeta_3&<&10^{-8},\nb\\
\zeta_4&<&6\times10^{-3},
\eqn
where
\bq
\lb{Gamma}
\Gamma \equiv 4\beta-\gamma-3-\frac{10}{3}\xi-\alpha_1+\frac{2}{3}\alpha_2-\frac{2}{3}\zeta_1-\frac{1}{3}\zeta_2.
\eq
In the following, we shall consider these experiment limits in various cases.

\subsection{$\sigma_1 = 0 = \sigma_2$}

In this case, from Eqs.(\ref{4.39}) and (\ref{4.40}) we find that
\bqn
\lb{4.43aa}
\beta&=&\frac{-a_1^3\gamma _1\varkappa^2+3 a_1^2 \gamma _1\varkappa+2 a_1 \gamma _1+a_1\varkappa+3}{4 a_1 \gamma _1+4}, \nb\\
\gamma&=&\frac{\varkappa(a_1^2 \gamma _1+a_2 a_1 \gamma _1+a_1)-a_2 \gamma _1}{a_1 \gamma _1+1},\nb\\
\alpha_1&=&\frac{4 \left(a_2-a_1\right) \gamma _1-4}{a_1 \gamma _1+1}\nb\\
&&-4 \varkappa  \left[a_2 \left(1-\frac{1}{a_1 \gamma _1+1}\right)+a_1-2\right],\nb\\
\alpha_2&=&\frac{\varkappa  \left[a_1^2 (3 \lambda -1)+a_1 (2-6 \lambda )+4 \lambda -2\right]-\lambda +1}{\lambda -1},\nb\\
\zeta_1&=&-\zeta_B=\frac{\left(a_1-1\right) a_1 (3 \lambda -1) \varkappa }{\lambda -1},\nb\\
\alpha_3&=& \xi = \zeta_2=\zeta_3=\zeta_4=0,
 \eqn
with
 \bqn
\lb{4.42aa}
\beta_0\left[a_1^2\varkappa \gamma _1+1\right] + 2\varkappa \left(a_1 \gamma
_1+1\right){}^2 = 0.
 \eqn
Then, from the experimental limit on   $\zeta_1$, we find that
\bq
\lb{a1C}
|a_1 - 1| \ll 1.
\eq
Expanding Eq.(\ref{4.43aa})  in terms of  $\epsilon \equiv a_1 -1$, we find
 \bqn
 \lb{4.44aaa}
\beta&=&1+\frac{\epsilon}{4}(1+\varpi)+{\cal O}(\epsilon^2),\nb\\
\gamma&=&1+\frac{1+\gamma_1+a_2\gamma_1}{1+\gamma_1}(1+\varpi)\epsilon+{\cal O}(\epsilon^2),\nb\\
\alpha_1&=&-4\frac{1+\gamma_1+a_2\gamma_1-\varpi(1+\gamma_1-a_2\gamma_1)}{1+\gamma_1}\epsilon+{\cal O}(\epsilon^2),\nb\\
\alpha_2&=&\varpi\epsilon+\frac{1-3\lambda}{1-\lambda}\epsilon^2+{\cal O}(\epsilon^3),\nb\\
\zeta_1&=&-\zeta_B=\frac{1-3\lambda}{1-\lambda}\epsilon,\nb\\
\alpha_3&=&\xi= \zeta_2 = \zeta_3=\zeta_4=0,
 \eqn
 where in writing the above expressions we had set  $\varkappa\equiv 1+\varpi\epsilon$. Thus, the experimental constraint on $\beta$ leads to
\bq
|a_1 -1 | < 10^{-5},\;\;\;\;
\varpi \simeq {\cal{O}}(1),
\eq
while     the constraints of $\gamma$ and $\alpha_1$ yield,
 \bq
\lb{CaseAa2}
a_2 = \cases{1+\gamma_1 +  {\cal{O}}\left(\epsilon\right), & $\gamma_1 \simeq -1$,\cr
 {\cal{O}}\left(1\right), & Otherwise. \cr}
\eq
On the other hand,  the constraints from  $\zeta_1$ and $\alpha_2$  further  require that $a_1$ must satisfy the constraints,
 \bqn
&& |a_1 - 1| < 10^{-7},\;\;\; \varpi \simeq {\cal{O}}(1),\nb\\
&& \lb{epsilon} \left|\frac{a_1 -1}{1- \lambda}\right| < 10^{-2},\;\;\;
\frac{(a_1 - 1)^2}{|1-\lambda|} < 10^{-7}.
 \eqn

It is interesting to note that the relativistic limit of the gravitational sector is
\bq
\lb{GRlimit}
(\gamma_1, \beta_0, \varkappa, \lambda)^{GR} = (-1, 0, 1, 1).
\eq
From Eqs.(\ref{4.42aa}) and (\ref{CaseAa2}) we find that for $(a_1, a_2) \simeq (1, 0)$, we indeed have  $(\gamma_1, \beta_0, \varkappa) = (-1,  0, 1)$. In fact, for 
\bq
\lb{ConditionAA}
a_1 = 1 = \varkappa,
\eq
Eq.(\ref{4.43aa}) reduces to
\bqn
\lb{PPNA}
&& \gamma = \beta = 1,\;\;\; \alpha_1 = \alpha_2 = \alpha_3 = \xi = 0,\nb\\
&& \zeta_1 =  \zeta_2 =  \zeta_3 =  \zeta_4 = \zeta_B = 0,\; (a_1 = 1),
\eqn
which are precisely the values obtained in GR. It should be noted that in the current case $\lambda$ cannot be taken exactly its relativistic value $\lambda_{GR} =1$,
in order to solve   the strong coupling problem by the mechanism proposed in  \cite{ZWWS,BPS} \footnote{See also the comments on this issue given in Footnote 1.}.

\subsection{$\sigma_{1} \not= 0, \; \sigma_2 = 0$}

In this case, we find that
 \bqn
 \lb{4.43}
\beta&=&\frac{-a_1^3\gamma _1\varkappa^2+3 a_1^2 \gamma _1\varkappa+2 a_1 \gamma _1+a_1\varkappa+3}{4 a_1 \gamma _1+4}\nb\\
&&+\frac{\gamma_1(a_1\varkappa-1)(1+a_1^2\gamma_1\varkappa)^2}{2\varkappa(1+a_1\gamma_1)^3}\sigma_1, \nb\\
\gamma&=&\frac{\varkappa(a_1^2 \gamma _1+a_2 a_1 \gamma _1+a_1)-a_2 \gamma _1}{a_1 \gamma _1+1},\nb\\
\alpha_1&=&\frac{4 \left(a_2-a_1\right) \gamma _1-4}{a_1 \gamma _1+1}\nb\\
&&-4 \varkappa  \left[a_2 \left(1-\frac{1}{a_1 \gamma _1+1}\right)+a_1-2\right],\nb\\
\alpha_2&=&\frac{\varkappa  \left[a_1^2 (3 \lambda -1)+a_1 (2-6 \lambda )+4 \lambda -2\right]-\lambda +1}{\lambda -1},\nb\\
\zeta_1&=&-\zeta_B=\frac{\left(a_1-1\right) a_1 (3 \lambda -1) \varkappa }{\lambda -1},\nb\\
\alpha_3&=& \xi = \zeta_2=\zeta_3=\zeta_4=0,
 \eqn
with $\beta_0$ being still given by Eq.(\ref{4.42aa}).
From the above expressions, it can be seen that $\sigma_1$ has contribution only to $\beta$.
Thus, the limit from  $\zeta_1$ still lead to the same constraint (\ref{a1C}). Then, we find that
\bqn
 \lb{4.44aaaa}
\beta&=&1+\frac{\epsilon}{4}(1+\varpi)  +\frac{\gamma_1(1+\varpi)}{8(1+\gamma_1)}\sigma_1\epsilon \nb\\
&&+{\cal O}\left(\frac{\epsilon^2}{(1+\gamma_1)^2}\right)+{\cal
O}(\epsilon^2).
 \eqn
Thus, the constraint from $\beta$ requires,
 \bqn
\lb{CaseBsigma1}
&& |a_1 -1 | < 10^{-5},\;\;\;\; \varpi \simeq {\cal{O}}(1),\nb\\
&& \sigma_1 = \cases{1+\gamma_1 +  {\cal{O}}\left(\epsilon\right), & $\gamma_1 \simeq -1$,\cr
 {\cal{O}}\left(1\right), & Otherwise. \cr}
\eqn
In addition,  the constraints from $\alpha_2$ and $\zeta_1$ further  require that $a_1$ must satisfy the constraints (\ref{epsilon}).

When $a_1 = \varkappa = 1$,   Eq.(\ref{4.43}) reduces exactly to    Eq.(\ref{PPNA}).

\subsection{$\sigma_1\sigma_2 \not= 0$}

 In review of the above analysis, let us consider the cases   $a_1 =1$ and $a_1 \not= 1$ separately.

 \subsubsection{$a_1 = 1$}

 Then, we find that
 \bqn
 \lb{a1result}
\beta&=&\frac{1}{4\varkappa[\gamma_1(\sigma_2-4)-4]^{3}}\bigg\{\gamma
_1 \sigma _2^3 \left[\gamma _1 \left(2 \gamma _1-1\right) \varkappa
+1\right]\nb\\
&&-4 \gamma _1 \sigma _2^2 \big[\varkappa  \left(\gamma
   _1 \left(2 \gamma _1+1\right) (\varkappa +3)-5\right)+1\big]\nb\\
   &&-8 \sigma _2 \big[\gamma _1 \sigma _1 \left(\gamma _1 \varkappa
   +1\right)^2\nb\\
   &&-2 \left(\gamma _1+1\right){}^2 \varkappa  \left(\gamma _1 (5 \varkappa +6)-1\right)\big]\nb\\
   &&-32 \gamma _1 \sigma _1
   (\varkappa -1) \left(\gamma _1 \varkappa +1\right)^2\nb\\
   &&+64 \left(\gamma _1+1\right)^2 \varkappa  \big[\gamma _1 ((\varkappa -3)
   \varkappa -2)-\varkappa -3\big]\bigg\},\nb\\
\gamma&=&-\frac{a_2 \gamma _1 \left(\sigma _2+4 \varkappa
-4\right)+4 \left(\gamma _1+1\right) \varkappa +\sigma _2}{\gamma _1
\left(\sigma
   _2-4\right)-4},\nb\\
\alpha_1&=&\frac{4 \gamma _1 \left(\sigma _2 \left(a_2+2 \varkappa
-1\right)+4 \left(a_2-1\right) (\varkappa -1)\right)}{\gamma _1
\left(\sigma
   _2-4\right)-4}\nb\\
   &&+\frac{4 \left(\sigma _2-4 \varkappa +4\right)}{\gamma _1 \left(\sigma _2-4\right)-4},\nb\\
\alpha_2&=&\frac{(3 \lambda -1) \sigma _2^2}{(\lambda -1) \varkappa
\left(\gamma _1 \left(\sigma
_2-4\right)-4\right){}^2}\nb\\
&&+\frac{\gamma _1 \sigma
   _2^2 \left(\gamma _1 (-\lambda +(4 \lambda -2) \varkappa +1)+6 \lambda -2\right)}{(\lambda -1) \left(\gamma _1 \left(\sigma
   _2-4\right)-4\right){}^2}\nb\\
   &&-\frac{8 \left(\gamma _1+1\right) (\varkappa -1) \left(\gamma _1 \left(\sigma
   _2-2\right)-2\right)}{\left(\gamma _1 \left(\sigma _2-4\right)-4\right){}^2},\nb\\
\zeta_1&=&-\zeta_B=\frac{(3 \lambda -1) \left(4 \left(\gamma
_1+1\right) \varkappa +\sigma _2\right) \left(\gamma _1 \sigma _2
\varkappa +\sigma
   _2\right)}{(\lambda -1) \varkappa  \left(\gamma _1 \left(\sigma _2-4\right)-4\right){}^2},\nb\\
\alpha_3&=&\xi= \zeta_2 = \zeta_3=\zeta_4=0, \eqn
with
 \bqn
 \lb{beta0C}
&& \gamma _1 \left(\sigma _2-8\right) \sigma _2+8 \varkappa  \left[\gamma _1 \sigma _2-2 \left(\gamma _1+1\right){}^2\right]\nb\\
&& ~~~~~~~~~~~~~~~~~~~ - 8(\gamma_1 \varkappa +1)\beta_0 = 0.
 \eqn
Considering the fact that   $(\gamma_1, \varkappa)^{GR} = (-1, 1)$, we  set
\bq
\lb{gamma1}
 \gamma_1=-1+\eta,~~~\varkappa=1+\varrho\eta,
\eq
where $\varrho \simeq {\cal{O}}(1)$. Then, expanding Eq.(\ref{a1result}) to the first-order of $\eta$, we obtain
 \bqn \lb{a1resultexpand}
\beta&=&1+\frac{3-\varrho}{4}\eta+{\cal O}(\eta^2),\nb\\
\gamma&=&1-a_2+\left[1+4\frac{a_2}{\sigma_2}(1-\varrho)\right]\eta+{\cal O}(\eta^2),\nb\\
\alpha_1&=&4a_2-4\left(1+4\frac{a_2}{\sigma_2}\right)(1-\varrho)\eta+{\cal O}(\eta^2),\nb\\
\alpha_2&=&\varrho\eta+\frac{1-3\lambda}{1-\lambda}(1-\varrho)^2\eta^2+{\cal O}(\eta^3),\nb\\
\zeta_1&=&-\zeta_B=\frac{1-3\lambda}{1-\lambda}(1-\varrho)\eta+{\cal O}(\eta^2),\nb\\
\alpha_3&=&\xi= \zeta_2 = \zeta_3=\zeta_4=0.
 \eqn
Following similar analysis given in the last two cases, it can
be shown that the experimental limits (\ref{4.47}) require
 \bqn
\lb{constraintD}
&& |\eta|<10^{-7},\;\;\;a_2<10^{-7},\;\;\; \left|\frac{a_2}{\sigma_2}\right|,\; \varrho  \leq {\cal O}(1), \nb\\
&&  \left|\frac{\gamma_1 +1}{1- \lambda}\right| <
10^{-2},\;\;\; \frac{(\gamma_1 + 1)^2}{|1-\lambda|} < 10^{-7}.
 \eqn


 \subsubsection{General $a_1$ and $\lambda=\frac{1}{3}$}

In this case, if  the coupling  constants  satisfy the relations,
\bqn
\lb{GR1}
a_1&=&-\frac{1}{\gamma_1}-a_2,\nb\\
\sigma_1&=&\frac{2\gamma_1a_2^2}
{(1+\gamma_1+2\gamma_1a_2+\gamma_1^2a_2^2)^2}\nb\\
&&\times \left\{ a_2 \left[2 a_2^2 \left(a_2^2-2a_2-1\right)+5a_2+2\right] \gamma _1^3 \right.\nb\\
&&\left.+\left[a_2^2 \left(5a_2^2-10a_2+1\right)+6a_2+1\right] \gamma _1^2\right.\nb\\
&&\left.+\left[a_2 \left(4 a_2^2-7a_2+2\right)+2\right]
\gamma_1+\left(a_2-1\right)^2\right\},\nb\\
\sigma_2&=&4 a_2 (a_2-1),\nb\\
\beta_0&=& \frac{2\gamma_1a_2}
{1+\gamma_1+2\gamma_1a_2+\gamma_1^2a_2^2} \left[2(\gamma_1+1)-2a_2
 \right.\nb\\
&&\left. +a_2\gamma_1\left(1-\gamma_1-4a_2+a_2^2\right)\right], \eqn
we immediately get the GR values of the PPN parameters, \bqn\lb{gr}
&&\varkappa=1,\;\; \beta=1=\gamma,\;\;
\alpha_1= \alpha_2=\alpha_3=0,\nb\\
&& \xi=\zeta_1=\zeta_2=\zeta_3=\zeta_4=\zeta_B=0.
\eqn

 \subsubsection{General $a_1$ and $\lambda$}

Seting 
\bq 
\lb{GR2} 
\sigma_2 = 4(1-a_1),\;\;\; \beta_0 =
-2(\gamma_1+1), 
\eq 
we can also obtain the GR values  of Eq.(\ref{gr}).

%
%
%
%

 \section {Post-Newtonian Approximations with projectability condition}
\renewcommand{\theequation}{6.\arabic{equation}} \setcounter{equation}{0}

In this section, we shall generalize the previous studies to the case with the projectablity condition
$N = N(t)$. In the following, we shall first give a brief review of this version of the HL theory, and then study its post-Newtonian
approximations. The total action can be written as \cite{HW},
\bqn
\lb{6.1}
S &=& \zeta^2\int dt d^{3}x N \sqrt{g} \Big({\cal{L}}_{K} -
{\cal{L}}_{{V}} +  {\cal{L}}_{{\varphi}} +  {\cal{L}}_{{A}} +  {\cal{L}}_{{\lambda}} \nb\\
& & ~~~~~~~~~~~~~~~~~~~~~~ \left. + {\zeta^{-2}} {\cal{L}}_{M} \right),
 \eqn
where ${\cal{L}}_{K}, \; {\cal{L}}_{\varphi}$ and ${\cal{L}}_{A}$ are given by Eq.(\ref{2.2}) with
$a_i = 0$,  but the potential    ${\cal{L}}_{{V}}$ now is given by
 \bqn \lb{6.2}
{\cal{L}}_{{V}} &=& \zeta^{2}g_{0}  + g_{1} R + \frac{1}{\zeta^{2}}
\left(g_{2}R^{2} +  g_{3}  R_{ij}R^{ij}\right)\nb\\
& & + \frac{1}{\zeta^{4}} \left(g_{4}R^{3} +  g_{5}  R\;
R_{ij}R^{ij}
+   g_{6}  R^{i}_{j} R^{j}_{k} R^{k}_{i} \right)\nb\\
& & + \frac{1}{\zeta^{4}} \Big[g_{7}(\nabla R)^{2} +  g_{8}
\left(\nabla_{i}R_{jk}\right) \left(\nabla^{i}R^{jk}\right)\Big],
~~~~
 \eqn
 where the coupling  constants $ g_{s}\, (s=0, 1, 2,\dots 8)$  are all dimensionless, and
 \bq
 \lb{6.3}
 \Lambda = \frac{1}{2} \zeta^{2}g_{0},
 \eq
 is the cosmological constant.

 In the IR, all the high-order derivative terms represented by $g_{2, ..., 8}$ are negligible, so in the rest of
 this section we shall set them to zero. In addition, with the same reasons as in the non-projectable case, here we also set $
 \Lambda = \Lambda_g = 0$. Moreover, using the U(1) gauge freedom, we choose the same gauge $\varphi = 0$, which uniquely fixes the gauge
 \cite{HW}. Then, the variation of the action (\ref{6.1}) with respect to $N, N_i, A, \varphi$ and $g_{ij}$, yield the Hamiltonian, momentum, $A$-, $\varphi$-  constraints
 and dynamical equations, respectively,  
 \bqn
 \lb{LowHamiltonain}
 && \int d^3x\sqrt{g}({\cal L}_K+ g_1 R =8\pi G\int d^3x\sqrt{g}J^t,\\
  \lb{LowMomentum}
&&  \nabla^j\pi_{ij}=8\pi GJ_i,\\
 \lb{LowA}
 && R=8\pi GJ_A,\\
 \lb{Lowvarphi}
&&  G^{ij}K_{ij}+(1-\lambda)\nabla^2K=8\pi GJ_\varphi,\\
 \lb{Lowdynamical}
 &&\frac{1}{\sqrt{g}}\left(\sqrt{g}\pi_{ij}\right)_{,t}+2K^l_iK_{lj}-2\lambda KK_{ij}\nb\\
 &&-\nabla^k\left(N_k\pi_{ij}-N_j\pi_{ik}-N_i\pi_{jk}\right)\nb\\
 &&-\frac{g_{ij}}{2}\left({\cal L}_K+{\cal L}_A\right)-g_1 G_{ij}\nb\\
 &&-AR_{ij}+\nabla_i\nabla_jA-g_{ij}\nabla^2A=8\pi G\tau_{ij},
 \eqn
where $G_{ij} [=  R_{ij} - Rg_{ij}/2]$ denotes the Einstein tensor.

Writing the metric $\gamma_{\mu\nu}$ in the forms (\ref{4.25}) and (\ref{4.26}), where $N, N_{i}, g_{ij}, A$ and $\varphi$ are
related to $\gamma_{\mu\nu}$ also through the universal coupling that we proposed in Section IV, given by Eqs.(\ref{eq8-1}) and (\ref{Pmetric}),
we find that in the present case there are also eleven independent PPN parameters, in terms which the metric $\gamma_{\mu\nu}$ takes the form
of Eq.(\ref{4.23}). Note that in writing $\gamma_{\mu\nu}$ in this form, we already fixed the gauge freedom by setting the anisotropic term $\chi_{,ij}$
in $\gamma_{ij}$ to zero. This uniquely fixes the gauge, as one can see from Eq.(\ref{4.22}).

Combining Eqs.(\ref{eq8-1}), (\ref{Pmetric}) and (\ref{4.25}), we  find
\bqn
\lb{S1F}
h_{00}=a_1\left[2(A_2+A_4) -a_1A_2^2\right] + {\cal O}(6),
\eqn
where we had expanded $A$ as
\bqn
\lb{S1A}
A=A_2+A_4 + {\cal O}(6),
\eqn
with $A_n$ being of order ${\cal O}(n)$.  Then,  comparing Eq.(\ref{S1F})  with Eq.(\ref{hmunu})  we obtain,
\bqn
\lb{S1R}
A_2=\frac{U}{a_1}.
\eqn
Thus, we have
\bqn
\lb{S1}
h_{00}=2U-U^2+2a_1A_4 + {\cal O}(6).
\eqn

Before proceeding further, we note that the Hamiltonian constraint in the projectable case becomes non-local. In addition,
when we consider the solar system tests, we implicitly assumed that we are considering a system large enough so that
$\eta_{\mu\nu}$ can be considered as the asymptotic form of $\gamma_{\mu\nu}$ [cf. Eq.(\ref{4.25})], yet small enough that
the deviation of the cosmological metric from  $\eta_{\mu\nu}$   is small for distances much smaller than
$r_{0} \equiv (M_{\bigodot}/a_0)^{1/2}  \simeq 10^{11}\; {\mbox{km}}$ \cite{Will}, where
$M_{\bigodot}$ denotes the mass of the solar system, and $a_0$ is the
current scale factor of the Universe.
Therefore, to consider the global Hamiltonian constraint, we need to know the space-time for  $ r > r_0$ and its corresponding
matter distributions. This is a complicated  question.  Fortunately, to study  the PPN parameters, we find that we do not need
to consider this global constraint.

With the above in mind, in the following we shall follow what we did in the last section and derive the PPN parameters in terms of the
coupling constants of the theory.

i) Momentum constraint to ${\cal O}(3)$: To this order, we find that  $J_i = -2\varkappa \rho v_i$, and Eq.(\ref{LowMomentum}) becomes
 \bqn
 \lb{6.31}
&& \Big[\frac{c}{2}+2d+\gamma-\lambda(c+3\gamma+2d)\nb\\
&& ~~ +1 - 3\lambda\Big]U_{,0i} -\frac{c}{2}\partial^2V_i
=2\varkappa \partial^2V_i,
 \nb
  \eqn where $c$ and $d$ are parameters appearing
in Eq.(\ref{hmunu}) with $b = 0$ in the current gauge. Therefore, we
obtain $c = -4\varkappa $, and
 \bqn
 \lb{6.32}
 \gamma+ 2d+ 1 - 3\lambda  -\lambda(3\gamma+2d - 4\varkappa) - 2\varkappa = 0.
 \eqn

ii) Dynamical equations to ${\cal O}(2)$: In this order, we find
$\tau_{ij} = 0$, and Eq.(\ref{Lowdynamical}) becomes
 \bqn
 \lb{6.34a}
\left(\frac{g_1a_2+1}{a_1}+g_1\gamma\right)\left(U_{,ij}-\delta_{ij}\partial^2U\right)=0,\nb
 \eqn
 from which we obtain
  \bq
  \lb{6.35}
\gamma=-\frac{g_1a_2+1}{g_1a_1}.
 \eq

iii)  $A$-constraint  to ${\cal O}(2)$: In this order, we find that  $J_A = 2a_1\rho$, and that Eq.(\ref{LowA}) becomes
 \bqn
 \lb{6.36}
 \left(\gamma+\frac{a_2}{a_1}\right)\partial^2U   = \varkappa a_1\partial^2U. \nb
 \eqn
Thus, we have
\bqn
\lb{6.36a}
\gamma=\varkappa a_1-\frac{a_2}{a_1}.
\eqn

iv)  $\varphi$-constraint  to ${\cal O}(3)$: In this order, we find that  $J_\varphi = a_1\rho_{,0}+(\rho v_k)_{,k}$ and Eq.(\ref{Lowvarphi}) becomes
 \bqn
 \lb{6.36bb}
&& (1-\lambda)\left(3\lambda+2d+\frac{3a_2}{a_1} - 4 \varkappa\right)\partial^2U_{,0}\nb\\
&& ~~~~~~~~~~~~~~~~~~~~~~ = 2\varkappa (a_1-1)\partial^2U_{,0},\nb
 \eqn
from which we obtain
 \bqn
\lb{6.36b} (1-\lambda)\left(3\lambda+2d+\frac{3a_2}{a_1}-
4\varkappa\right)= 2\varkappa(a_1-1).
 \eqn
Similar to the non-projectable case, this equation is also not independent, and can be obtained from   Eqs.(\ref{6.32}), (\ref{6.35}) and  (\ref{6.36a}).

v)  The $A$-constraint  and the trace of dynamical equations to ${\cal O}(4)$:  To this order, we find that the trace of $\tau_{ij}$ to ${\cal O}(4)$ is $\delta_{ij}\tau_{ij}=\rho v^2+3p$,
and the dynamical equations to ${\cal O}(4)$ yield,
 \bqn
 \lb{6.38}
&&-\frac{1-3\lambda}{2}\left(2d-4+3\gamma+3\frac{a_2}{a_1}\right)\partial^2(\mathfrak
{A}+\mathfrak{B}-\Phi_1)\nb\\
&&-\frac{2}{a_1}\partial^2U+\frac{4}{a_1}\left(\gamma+\frac{a_2}{a_1}\right)\partial^2\Phi_2-2\partial^2A_4\nb\\
&&-\left(\frac{\gamma}{a_1}+\frac{a_2}{a_1^2}\right)\partial^2U^2+\frac{2}{a_1}\left(\gamma+\frac{a_2}{a_1}\right)\partial^2\Phi_2\nb\\
&=&\varkappa \partial^2\Big[2g_1a_1U+2(1-3g_1a_2)\Phi_2+2g_1a_1(\Phi_1+\Phi_3)\nb\\
&&-2(1+g_1a_2)(\Phi_1+3\Phi_4)\Big].
 \eqn
Note that in writing the above equation, we had used Eq.(\ref{LowA}) to replace $R$. Since all the quantities appearing in the above equation are known upto
${\cal O}(4)$, except for the fourth-order term $A_4$. Thus, first solving the above equation for $A_4$, and  then substituting it into Eq.(\ref{S1F}), we finally
obtain,
\bqn
\lb{6.39}
h_{00}&=&-\left(1+\frac{a_2}{a_1}+\gamma\right)U^2\nb\\
&&-\frac{1-3\lambda}{2}[3a_2 +a_1(2d+3\gamma - 4\varkappa)](\mathfrak
{A}+\mathfrak{B})\nb\\
&&+\frac{a_1}{2}\Big\{(1-3\lambda)\left(2d+3\gamma+3\frac{a_2}{a_1}-4\varkappa\right)\nb\\
&&+4\varkappa+4g_1\varkappa(a_2-a_1)\Big\}\Phi_1\nb\\
&&+2\left(2\gamma+2\frac{a_2}{a_1}+3\varkappa a_1a_2g_1\right)\Phi_2\nb\\
&&-2a_1^2\varkappa g_1\Phi_3+6\varkappa a_1(1+a_2g_1)\Phi_4.
\eqn
Comparing it with Eq.(\ref{4.23}) and considering the fact
$\gamma_{00} = - 1 + h_{00}$, we obtain
 \bqn
\lb{6.40}
\beta&=&\frac{1 + \varkappa a_1}{2},\nb\\
\gamma&=&\varkappa a_1-\frac{a_2}{a_1},\nb\\
\alpha_1&=&4\left[(\varkappa-1) - (a_1-1)\varkappa +\frac{a_2}{a_1}\right], \nb\\
\alpha_2&=&  (\varkappa -1) + \left(\frac{1-3\lambda}{1-\lambda}\right)(a_1-1)^2\varkappa,\nb\\
\alpha_3&=&\xi=0,\nb\\
\zeta_1&=&-\zeta_B= \left(\frac{1-3\lambda}{1-\lambda}\right)a_1(a_1-1)\varkappa, \nb\\
\zeta_2&=&\zeta_3=\zeta_4=0,
 \eqn
where
\bqn
\lb{6.41}
a_1  &=& \pm \frac{1}{\sqrt{-\varkappa g_1}},\nb\\
d &=& \varkappa\frac{2- a_1-\lambda(4-3a_1)}{2(1-\lambda)}.
 \eqn
Thus,  the experimental limits of $\beta$, $\gamma$ and $\alpha_1$  require
 \bq
\lb{a1a2} |a_1 -1| < 10^{-5},\;\;\; |a_2| <
10^{-5},\;\;\;|\varkappa-1|<10^{-5},
 \eq
while the limit of  $\alpha_2$ leads to
 \bq
 \lb{alpha2}
(\varkappa-1),\;\;
 \left|\frac{(a_1-1)^2}{1-\lambda}\right| < 10^{-7}.
 \eq
Once the above two conditions are satisfied, it can be seen that all the rest of experimental limits given in Eq.(\ref{4.47}) are satisfied identically.

It is interesting to note that with the constraint (\ref{a1a2}), we must
choose the ``+'' sign in Eq.(\ref{6.41}), from which we find that
Eq.(\ref{a1a2}) yields,
\bq
\lb{g1}
|g_1 + 1| < 10^{-5}.
\eq
It is remarkable to note that the relativistic limit of this version of the HL theory is
\bq
\lb{GRLimit}
( \lambda, g_1)^{GR}  = (1, -1).
\eq

In addition, when $a_2 = 0$ the prescription of Eq.(\ref{eq8-1}) exactly reduces  to the one proposed in \cite{Lin:2012bs}, for which the
line element,
\bqn
\lb{lineE}
ds^2 &=& \gamma_{\mu\nu}dx^{\mu}dx^{\nu} \nb\\
&=& - {\tilde{N}}^2dt^{2} + \tilde{g}_{ij}\left(dx^i + \tilde{N}^idt\right)\left(dx^j + \tilde{N}^j dt\right),\nb
\eqn
is invariant not only under the foliation-preserving transformations,  Diff($M, \; {\cal{F}}$),  but also under the local U(1) transformations.
Since the constraints  for  the nonprojectable case does not impose any limit on the value of $a_2$, combing this  with the limit of Eq.(\ref{a1a2})
given for the projectable case, we can see that the requirement that $ds^2$  be invariant with respect to the enlarged symmetry (\ref{symmetry}) is consistent
with the solar system tests and fix the metric to the form,
\bqn
\lb{lineEb}
ds^2 
&=& - {\cal{N}}^2dt^{2} + {g}_{ij}\left(dx^i + {\cal{N}}^idt\right)\left(dx^j + {\cal{N}}^j dt\right), ~~~~~~~
\eqn
with \cite{Lin:2012bs}
\bqn
\lb{NNi}
{\cal{N}} &=& N\left(1 + \frac{\upsilon}{c^2} \frac{A - \cal{A}}{N}\right),\nb\\
{\cal{N}}^i &=& N^i + N\nabla^i\varphi,
\eqn
where $\upsilon [\equiv a_1]$ is the parameter introduced in \cite{Lin:2012bs}. The considerations of the solar system tests for the Eddington-Robertson-Schiff parameters
$\gamma$ and $\beta$ in the spherical vacuum case led to the constraint \cite{Lin:2012bs}
\bq
\lb{upsilon}
|\upsilon-1|<10^{-5},
\eq
which is precisely the one given by Eq.(\ref{a1a2}).

 \section{Conclusions}
\renewcommand{\theequation}{7.\arabic{equation}} \setcounter{equation}{0}

In this paper, we have studied the post-Newtonian approximations in the
framework of the HL theory of gravity with the enlarged symmetry
(\ref{symmetry}) for  the projectable $N = N(t)$ as well as the
nonprojectable $N= N(t, x)$ cases. Contrary to the previous works in
which only two PPN parameters were calculated~\cite{Lin:2012bs,LW}, we
obtain {\it all} PPN parameters in both cases. We then found parameter
regions in which the theory passes {\it all} solar system tests. This
had never been possible before, in either projectable or non-projectable 
case.

To study these approximations, one needs first to specify the coupling
between the HL gravity and matter, which is still an open question in
the HL theory \cite{reviews}. Motivated by our previous studies of the
post-Newtonian approximations in the static spherical vacuum case
\cite{Lin:2012bs}, we have considered the scalar-tensor extensions
[cf. Appendix C], and found a universal coupling in the IR, given by
Eqs.(\ref{eq8-1}) and (\ref{eu7}), which reduces to the one proposed in
\cite{Lin:2012bs} when $a_2 = 0$, for which the metric takes the form of
Eq.(\ref{lineEb}). This metric is invariant not only under the
foliation-preserving  transformations,  Diff($M, \; {\cal{F}}$),  but
also under the local U(1) transformations. Although this universal
coupling was initially done for the projectable case, in this paper we
have also generalized it to the non-projectable case.

Before studying the post-Newtonian approximations using the above
universal coupling, we have considered the most general version of the
HL theory without the projectability condition [cf. Section II], in
which the term ${\cal{L}}_{S}$ given by Eq.(\ref{2.2}) has been added to
the action (\ref{action}). As shown explicitly in Section III, because
of the presence of the $\sigma_2$ term, spin-0 gravitons generically
appear, although they are stable in a large region of the parameters
space. This is different from the case without the ${\cal{L}}_{S}$ term,
considered in \cite{ZWWS}, in which it was shown that spin-0 gravitons
were eliminated by the extra U(1) symmetry, similar to the projectable
case \cite{HMT,WW,daSilva,HW}.

With the above in mind, we have studied the post-Newtonian
approximations first in the non-projectable case [Section V]. After
tedious calculations and analysis, we have finally shown that all the
solar system tests carried out so far, represented by Eq.(\ref{4.47}),
are satisfied by properly choosing the coupling constants of the
theory. In particular, with specific choices of the independent coupling
constants: (i) (\ref{ConditionAA}) with $\sigma_2 = 0$ and a general
$\lambda$, or (ii) (\ref{GR1}) with  $\sigma_2 \not= 0$ and $\lambda=1/3$,  
or (iii) (\ref{GR2})  with $\sigma_2 \not= 0$
and a  general
$\lambda$,  the PPN parameters are given  by
\bqn
\lb{7.1}
&& \gamma = \beta = 1,\;\;\;
\alpha_1 =  \alpha_2 =  \alpha_3 =  \xi = 0, \nb\\
&&  \zeta_1 = \zeta_2 = \zeta_3 = \zeta_4 = \zeta_{B} = 0,
\eqn
which are precisely the results obtained in GR. Interestingly, the last
choice (\ref{GR2}) includes the minimal coupling to matter as a special
case:
\bq
a_1=a_2=0, \;\;
\sigma_2=4, \;\;
 \beta_0 = -2(1+\gamma_1),\; [N = N(t, x)].
\eq
Note that this was made possible by the inclusion of the new term
$\sigma_2\ne 0$. With the above conditions,   the  scalar gravitons are also stable for
$\lambda>1$ or $ \lambda<1/3$.


Similar considerations have been carried out for the projectable case in
Section VI, and shown that the same values of the PPN parameters given
by Eq.(\ref{7.1}) can be also obtained
with the choice
\bq
\lb{7.2}
a_1=1, \;\;
a_2=0, \;\;
g_1=-1,\;\; [N = N(t)].
\eq
This excludes the minimal coupling to matter in the projectable case, as already shown in \cite{Lin:2012bs}. 

We also note that the GR limit of the PPN parameters in both of the projectable and non-projectable cases imposes no constraint on  $\lambda$, appearing in the kinetic part ${\cal{L}}_{K}$ of the action [cf. Eq.(\ref{2.2})]. This is in contrast to the other versions of the HL theory \cite{reviews}, in particular to the healthy extension \cite{BPS}, in which it was found that the solar system tests require $| 1 - \lambda| < 10^{-7}$.
As a result, the condition for the energy scale $M_*$ of the theory
\bq
\lb{Masslimit}
M_{*} \leq | 1 - \lambda|^{1/2} M_{pl},
\eq
leads to $M_{*}  \leq 10^{15-16}$ GeV, where $M_{pl}$ denotes the Planck mass. Therefore, in the case with the extra U(1) symmetry, the corresponding energy scale $M_{*}$ is not constrained
 by the solar system tests, and in principle can be taken any value.

Finally, it is worthwhile investigating cosmology in the theory with the
proper matter coupling taken into account. In the prescription proposed
in the present paper, the geometry on which matter propagates is
not given by $g_{\mu\nu}$, constructed only from  the ADM   
variables ($N, N^i, g_{ij}$) \cite{ADM},
\bqn
\lb{PmetricA}
\left(g_{\mu\nu}\right) &\equiv& \left(\matrix{-{N}^2 + {N}^{i}{N}_{i} & {N}_{i}\cr
{N}_{i} & {g}_{ij}\cr}\right),
\eqn
but by $\gamma_{\mu\nu}$ defined by Eq,(\ref{Pmetric}), which is
a combination of the fundamental HL variables ($N, N^i, g_{ij}, A, \varphi$) through Eqs.(\ref{eq8-1})-(\ref{eq8-2}). 
In addition, the  HL gravitational field couples {\em universally} to matter through 
 [cf.  (\ref{eu7})], 
\bq
\lb{eu7a}
S_{m} = \int{\sqrt{-\gamma} d^4x \tilde{\cal{L}}_{m}\left(\gamma_{\mu\nu}; \psi_{n}\right)},
\eq
 in which it is   $\gamma_{\mu\nu}$ that mininally 
couples to matter, instead of $g_{\mu\nu}$, as that given, for example,  in general relativity.
Yet, in the nonprojectable case even in the vacuum case
the spatial 3-dimensional curvature $R$ is not necessarily a constant, because of the presence of 
terms represetned by $\sigma_1$ and $\sigma_2$, introduced in this paper, as one can see from Eq.(\ref{A}). 
With all these changes, it is expected that cosmology in the 
framework of the HL theory proposed in this paper will be different from the ones studied previously
even all with the extra U(1) symmetry. These include  the spatial
flatness issue \cite{HW,ZWWS}. On the other hand, it is  interesting to note that in the HL gravity 
(with or without the extra $U(1)$), the horizon problem may be solved without
inflation~\cite{Mukohyama:2009gg}. Details of the HL cosmology in the framwork  proposed in this paper will
be studied in the future publications.

\section*{\bf Acknowledgements}

This work is supported in part by DOE  Grant, DE-FG02-10ER41692 (AW);  WPI Initiative, MEXT, Japan (SM),
Grant-in-Aid for Scientific Research 24540256 and 21111006 (SM);  NSFC No. 11375153 (AW), No. 11173021 (AW),
NSFC No. 11178018 (KL),  No. 11075224 (KL),
No. 11047008 (TZ), No. 11105120 (TZ), No. 11205133 (TZ);  Ci\^encia Sem Fronteiras,
No. 004/2013 - DRI/CAPES (AW), and  FAPESP No. 2012/08934-0 (KL).

\section*{Appendix A: Another derivation of the non-projectable HL gravity with U(1) symmetry}
\renewcommand{\theequation}{A.\arabic{equation}} \setcounter{equation}{0}

Considering the local U(1) transformations (\ref{A.0d}), we find that ${\cal{N}}_{i}$ defined by
\bqn
\lb{F0}
{\cal{N}}_{i} = N_{i} + N\nabla_i\varphi,
\eqn
is gauge-invariant, $\delta_{\alpha}{\cal{N}}_{i} = 0$, and has the dimension  $[{\cal{N}}_i] = 2$. Then, the quantity $\tilde{K}_{ij}$ defined by
\bqn
\lb{F0b}
\tilde{K}_{ij}&\equiv& \frac{1}{2N}\left(-\dot{g}_{ij} + \nabla_{i}{\cal{N}}_{j} + \nabla_{j}{\cal{N}}_{i}\right)\nb\\
&=& K_{ij}+\nabla_i\nabla_j\varphi+a_{(i}\nabla_{j)}\varphi,
\eqn
is also gauge-invariant under the  U(1) transformations. In addition, $\tilde{K}_{ij}$ has the same dimension as ${K}_{ij}$, i.e., $[\tilde{K}_{ij}] = 3$, from which one can see that
the quantity $ \tilde{\cal L}_K$,
\bq
\lb{F0c}
 \tilde{\cal L}_K \equiv \tilde{K}^{ij}\tilde{K}_{ij}-\lambda\tilde{K}^2,
 \eq
has dimension 6. On the other hand, $\sigma$ defined in Eq.(\ref{2.6}) is also gauge-invariant, and has the dimension $[\sigma] = 4$. Thus, the quantity
\bq
\lb{F0d}
\tilde{\cal{L}}_{S} \equiv \sigma\left(Z_A+2\Lambda_g-R\right),
\eq
has  dimension 6 and is gauge-invariant under the  U(1) transformations, where $Z_{A} \equiv \sigma_1 a^ia_i + \sigma_2 a^{i}_{\;\; i}$. Then, the
action
\bqn
  \lb{F1}
S_g&=&\zeta^2\int
dtd^3xN\sqrt{g}\big[\tilde{\cal L}_K-\gamma_1R-2\Lambda\nb\\
&&+\beta a^ia_i+\sigma\left(Z_A+2\Lambda_g-R\right)- {\cal
L}_{z>1}\big],
\eqn
is gauge-invariant under the  U(1) transformations (\ref{A.0d}), where $  {\cal L}_{z>1}$ denotes the potential part that contains spatial derivative operators higher than second-order.
To show that the above action is equal to that given by Eq.(\ref{action}), we first note that,
 \bqn\lb{F3}
&&\frac{1}{\sqrt{g}}\left[\partial_t\left(\sqrt{g}\right)-\partial_i\left(\sqrt{g}N^i\right)\right]=NK,\nb\\
&&\frac{1}{\sqrt{g}}\left[\partial_t\left(\sqrt{g}R\right)-\partial_i\left(\sqrt{g}N^iR\right)\right]=-2NG^{ij}K_{ij}\nb\\
&&~~~~~~~~+2\left[\nabla^i\nabla^j\left(NK_{ij}\right)-\nabla^2\left(NK\right)\right].
 \eqn
Then, we find,
 \bqn\lb{F4}
 &&\int dtd^3xN\sqrt{g}\sigma\nb\\
 &&~~=\int dtd^3x\sqrt{g}\Bigg[A-\left(\partial_t-N^i\partial_i\right)\varphi\nb\\
 &&~~~~~~~~~~~~~~~~ - \frac{1}{2}Ng^{ij}\left(\partial_i\varphi\right)\left(\partial_j\varphi\right)\Bigg]\nb\\
 &&~~=\int dtd^3x\sqrt{g}\left\{A+2N\left[K+\frac{1}{2}\left(\nabla^i+a^i\right)\nabla_i\varphi\right]\right\},\nb\\
 &&\int dtd^3xN\sqrt{g}R\sigma \nb\\
 &&~~=\int dtd^3x\sqrt{g}R\big[A-\left(\partial_t-N^i\partial_i\right)\varphi\nb\\
 && ~~~~~~~~~~~~~~~~ - \frac{1}{2}Ng^{ij}\left(\partial_i\varphi\right)\left(\partial_j\varphi\right)\big]\nb\\
 &&~~=\int dtd^3x\sqrt{g}\Big\{AR+2N\Big[-2G^{ij}K_{ij}\nb\\
 &&~~~~~~~~+2\left(\nabla^i+2a^i\right)\nabla^j\left(K_{ij}-g_{ij}K\right)\nb\\
 &&~~~~~~~~+2\left(a^i a^j+a^{ij}\right)\left(K_{ij}-g_{ij}K\right)\nb\\
 &&~~~~~~~~+\frac{1}{2}\left(\nabla^i R\right)\left(\nabla_i \varphi\right)+\frac{R}{2}\left(\nabla^i+a^i\right)\nabla_i \varphi \Big]\Big\},\nb\\
 &&\int dtd^3xN\sqrt{g}Z_A\sigma\nb\\
 &&~~=\int dtd^3x\sqrt{g}Z_A\big[A-\left(\partial_t-N^i\partial_i\right)\varphi\nb\\
 &&~~~~~~~~-g^{ij}\frac{N}{2}\left(\partial_i\varphi\right)\left(\partial_j\varphi\right)\big]\nb\\
 &&~~=\int dtd^3x\sqrt{g}\Bigg\{AZ_A\nb\\
 && ~~~~~~~~ +NZ_A\left[K+\frac{1}{2}\left(\nabla^i+a^i\right)\nabla_i\varphi\right]\nb\\
 &&~~~~~~~~+2N\left[\partial_\bot Z_A+\frac{1}{2}\left(\nabla^iZ_A\right)\left(\nabla_i\varphi\right)\right]\Bigg\},\nb\\
 &&\int dtd^3xN\sqrt{g}\tilde{\cal L}_K=\int dtd^3xN\sqrt{g}\left(\tilde{K}^{ij}\tilde{K}_{ij}-\lambda\tilde{K}^2\right)\nb\\
 &&~~=\int dtd^3xN\sqrt{g}\Bigg\{\left(K^{ij}K_{ij}-\lambda K^2\right)+R\left(\sigma-\frac{A}{N}\right)\nb\\
 &&~~~~~~~~+2(1-\lambda)\left(\nabla_i+a_i\right)\nabla^jK\nb\\
 &&~~~~~~~~+(1-\lambda)\left(\nabla_i+a_i\right)\nabla^j\left(\nabla_j+a_j\right)\nabla^j\varphi\nb\\
 &&~~~~~~~~+G^{ij}\left[2K_{ij}+\left(\nabla_i+a_i\right)\nabla_j\varphi\right]+\Delta {\cal
 L}_K\Bigg\},
 \eqn
where
 \bqn\lb{F5}
 \Delta{\cal
 L}_K&=&2\left(K^{ij}-g^{ij}K\right)a_i\nabla_j\varphi+\left(\nabla^ia^j\right)\left(\nabla_i\varphi\right)\left(\nabla_j\varphi\right)\nb\\
 &&-\left(\nabla_ia^i\right)\left(\nabla^j\varphi\right)\left(\nabla_j\varphi\right)
 +\frac{\varphi}{2}\bigg[a_i\left(\nabla^ja^i\right)\left(\nabla_j\varphi\right)\nb\\
 &&-a^j\left(\nabla_ia^i\right)\left(\nabla_j\varphi\right)+a^ia_i\nabla^2\varphi-a^ia^j\nabla_i\nabla_j\varphi\bigg].\nb\\
 \eqn
Inserting the above into Eq.(\ref{F1}), we obtain,
 \bqn\lb{F6}
 S_g&=&\zeta^2\int dtd^3xN\sqrt{g}\Big\{{\cal L}_K-\gamma_1R-2\Lambda\nb\\
&&+\beta a^ia_i+\frac{A}{N}\left(Z_A+2\Lambda_g-R\right)- {\cal L}_{z>1}\nb\\
&&+\varphi\left[\partial_\bot Z_A+\frac{1}{2}\left(\nabla^i\varphi\right)\left(\nabla_iZ_A\right)\right]\nb\\
&&+(1-\lambda)\big\{2K\left(\nabla_i+a_i\right)\nabla^j\varphi\nb\\
&&+\left[\left(\nabla_i+a_i\right)\nabla^j\varphi\right]^2\big\}+\Bigg[g^{ij}\left(\Lambda_g+\frac{Z_A}{2}\right)\nb\\
&&+G^{ij}\Bigg]\varphi\big[2K_{ij}+\left(\nabla_i+a_i\right)\nabla_j\varphi\big]\nb\\
&&+2\left(K^{ij}-g^{ij}K\right)a_i\nabla_j\varphi+\left(\nabla^ia^j\right)\left(\nabla_i\varphi\right)\left(\nabla_j\varphi\right)\nb\\
 &&-\left(\nabla_ia^i\right)\left(\nabla^j\varphi\right)\left(\nabla_j\varphi\right)
 +\frac{\varphi}{2}\bigg[a_i\left(\nabla^ja^i\right)\left(\nabla_j\varphi\right)\nb\\
 &&-a^j\left(\nabla_ia^i\right)\left(\nabla_j\varphi\right)+a^ia_i\nabla^2\varphi-a^ia^j\nabla_i\nabla_j\varphi\bigg],\nb\\
 \eqn
which is the same action as that given by Eq.(\ref{action}).

\section*{Appendix B: $F_V,\;F_\varphi,\;F_\lambda,\;F_{ij},\;F_{ij}^a\;$ and $F_{ij}^\varphi$}
\renewcommand{\theequation}{B.\arabic{equation}} \setcounter{equation}{0}

  $F_V,\;F_\varphi $ and $F_\lambda$, defined in Eq.(\ref{hami}),  are given by, 
\bqn\label{a1}
F_V &=&  \beta_0 ( 2 a_i^i + a_i a^i) - \frac{\beta_1}{\zeta^2} \Bigg[3 (a_i a^i)^2 + 4 \nabla_i (a_k a^k a^i)\Bigg]\nb\\
    &&  +\frac{\beta_2}{\zeta^2}\Bigg[ (a_i^i)^2 + \frac{2}{N} \nabla^2 (N a_k^k)\Bigg]\nb\\
    && - \frac{\beta_3}{\zeta^2}\Bigg[(a_i a^i) a_j^j + 2 \nabla_i (a_j^j a^i) - \frac{1}{N} \nabla^2 (N a_i a^i)\Bigg]\nb\\
    &&+ \frac{\beta_4}{\zeta^2}\Bigg[a_{ij} a^{ij} + \frac{2}{N} \nabla_j \nabla_i (N a^{ij})\Bigg]\nb\\
      && - \frac{\beta_5}{\zeta^2}\Bigg[R (a_i a^i) + 2 \nabla_i (R a^i)\Bigg]\nb\\
      &&- \frac{\beta_6}{\zeta^2}\Bigg[a_i a_j R^{ij} + 2\nabla_i (a_j R^{ij})
      \Bigg]\nb\\
      && +  \frac{\beta_7}{\zeta^2}\Bigg[ R a^i_i + \frac{1}{N} \nabla^2 (NR)\Bigg]\nb\\
      &&+ \frac{\beta_8}{\zeta^4}\Bigg[(\Delta a^i)^2 - \frac{2}{N} \nabla^i [\Delta (N \Delta a_i)]\Bigg],
\eqn \bqn\label{a2}
F_\varphi &=& -  {\cal{G}}^{ij}\nabla_i \varphi \nabla_j \varphi, - \frac{2}{N} \hat{{\cal{G}}}^{ijkl} \nabla_l (N K_{ij} \nabla_k \varphi),\nb\\
        &&  - \frac{4}{3}\hat{{\cal{G}}}^{ijkl} \nabla_l (\nabla_k \varphi \nabla_i \nabla_j \varphi)\nb\\
         &&- \frac{5}{3}\hat{{\cal{G}}}^{ijkl} \Big[(a_i \nabla_j \varphi) (a_k \nabla_l \varphi)+\nabla_i(a_k\nabla_j\varphi\nabla_l\varphi)\nb\\
         && +\nabla_k(a_i\nabla_j\varphi\nabla_l\varphi)\Big]\nb\\
         &&+ \frac{2}{3}\hat{{\cal{G}}}^{ijkl}\Big[a_{ik} \nabla_j \varphi \nabla_l \varphi + \frac{1}{N} \nabla_i\nabla_k (N\nabla_j\varphi\nabla_l\varphi)\Big],\nb\\
\eqn \bqn\lb{a3a}
F_\lambda &=& (1-\lambda) \Bigg\{(\nabla^2 \varphi + a_i \nabla^i \varphi)^2 - \frac{2}{N} \nabla_i (NK\nabla^i \varphi)\nb\\
           && - \frac{2}{N} \nabla_i \Big[N (\nabla^2 \varphi + a_i \nabla^i \varphi) \nabla^i \varphi\Big]\Bigg\}\label{a3}.
\eqn

 $\left(F_n\right)_{ij}$, $\left(F^{a}_{s}\right)_{ij}$ and $\left(F^{\varphi}_{q}\right)_{ij}$, defined in Eq.(\ref{tauij}),  are given, respectively, by
\bqn\lb{a4}
(F_0)_{ij} &=& -\frac{1}{2}g_{ij},\nb\\
(F_1)_{ij} &=& R_{ij}-\frac{1}{2}Rg_{ij}+\frac{1}{N}(g_{ij}\nabla^2 N-\nabla_j\nabla_i N),\nb\\
(F_2)_{ij} &=& -\frac{1}{2}g_{ij}R^2+2RR_{ij}\nb\\
             &&  +\frac{2}{N}\left[g_{ij}\nabla^2(NR)-\nabla_j\nabla_i(NR)\right],\nb\\
(F_3)_{ij} &=& -\frac{1}{2}g_{ij}R_{mn}R^{mn}+2R_{ik}R^k_{j}\nb\\
             &&  - \frac{1}{N}\Big[2\nabla_k\nabla_{(i}(NR_{j)}^k)\nb\\
             &&  -\nabla^2(NR_{ij})- g_{ij}\nabla_m\nabla_n(NR^{mn})\Big], \nb\\
(F_4)_{ij} &=&  -\frac{1}{2}g_{ij}R^3+3R^2R_{ij}\nb\\
             &&   +\frac{3}{N}\Big(g_{ij}\nabla^2-\nabla_j\nabla_i\Big)(NR^2),\nb\\
(F_5)_{ij} &=& -\frac{1}{2}g_{ij}RR_{mn}R^{mn}\nb\\
             &&+R_{ij}R_{mn}R^{mn}+2RR_{ik}R^k_{j}\nb\\
             &&  +\frac{1}{N}\Big[g_{ij}\nabla^2(NR_{mn}R^{mn})\nb\\
             &&-\nabla_j\nabla_i(NR_{mn}R^{mn})\nb\\
             &&  +\nabla^2(NRR_{ij})+g_{ij}\nabla_m\nabla_n(NRR^{mn})\nb\\
             &&  -2\nabla_m\nabla_{(i}(R^m_{j)}NR)\Big], \nb\\
(F_6)_{ij} &=& -\frac{1}{2}g_{ij}R^m_nR^n_lR^l_m+3R_{mn}R_{mi}R_{nj}\nb\\
             &&  +\frac{3}{2N}\Big[g_{ij}\nabla_m\nabla_n(NR^m_aR^{na}) \nb\\
             &&+ \nabla^2(NR_{mi}R^m_j)
              -2\nabla_m\nabla_{(i}(NR_{j)n}R^{mn})\Big], \nb\\
(F_7)_{ij} &=& -\frac{1}{2}g_{ij}R\nabla^2R+R_{ij}\nabla^2R+R\nabla_i\nabla_j R\nb\\
             &&  +\frac{1}{N}\Big[g_{ij}\nabla^2(N\nabla^2R)-\nabla_j\nabla_i(N\nabla^2R)\nb\\
             &&+R_{ij}\nabla^2(NR)
              +g_{ij}\nabla^4(NR)-\nabla_j\nabla_i (\nabla^2 (NR))\nb\\
             &&  - \nabla_{(j}(NR\nabla_{i)}R)+\frac{1}{2}g_{ij}\nabla_k(NR\nabla^kR)\Big], \nb\\
(F_8)_{ij} &=& -\frac{1}{2}g_{ij}(\nabla_mR_{nl})^2 + 2 \nabla^mR^n_i\nabla_mR_{nj}\nb\\
             &&  +\nabla_iR^{mn}\nabla_jR_{mn}+\frac{1}{N}\Big[2 \nabla_n\nabla_{(i}\nabla_m(N\nabla^mR^n_{j)})\nb\\
             &&  -\nabla^2\nabla_m(N\nabla^mR_{ij})-g_{ij}\nabla_n\nabla_p\nabla_m(N\nabla^mR^{np})\nb\\
             &&  -2\nabla_m(NR_{l(i}\nabla^mR^l_{j)})-2\nabla_n(NR_{l(i}\nabla_{j)}R^{nl})\nb\\
             &&  +2\nabla_k(NR^k_l\nabla_{(i}R^l_{j)})\Big], \nb\\
(F_9)_{ij} &=& -\frac{1}{2} g_{ij} a_k G^k+\frac{1}{2} \Big[a^k R_{k((j} \nabla_{i)} R + a_{(i} R_{)jk} \nabla^k R\Big]\nb\\
           &&-a_kR_{mi}\nabla_jR^{mk}-a^kR_{in}\nabla^nR_{jk}-a_iR^{km}\nabla_mR_{kj}\nb\\
           &&-\frac{3}{8}a_{(i}R\nabla_{j)}R+\frac{3}{8}\Bigg\{R\nabla_k(Na^k)R_{ij}\nb\\
           &&+g_{ij}\nabla^2\Big[R\nabla_k(Na^k)\Big]-\nabla_i\nabla_j\Big[R\nabla_k(Na^k)\Big]\Bigg\}\nb\\
           &&+\frac{1}{4N} \Bigg\{- \frac{1}{2}\nabla^m \Big[\nabla_i (Na_j\nabla_m R+Na_m\nabla_j R)\nb\\
           &&+\nabla_j (Na_i\nabla_m R+Na_m\nabla_i R)\Big]\nb\\
           &&+\nabla^2 (N a_{(i}\nabla_{j)}R)+g_{ij} \nabla^m\nabla^n (Na_m\nabla_nR)\nb\\
           && +\nabla^m\Big[\nabla_i(Na_k\nabla_jR^k_m+Na_k\nabla_mR^k_j)\nb\\
           &&+\nabla_j(Na_k\nabla_iR^k_m+Na_k\nabla_mR^k_i)\Big]\nb\\
           &&-2\nabla^2(Na_k\nabla_{(i} R^k_{j)})-2g_{ij}\nabla^m\nabla^n(Na_k\nabla_{(n}R_{m)}^k)\nb\\
           &&- \nabla^m \Big[\nabla_i\nabla_p(Na_jR_m^p+Na_mR_j^p)\nb\\
           &&+\nabla_j\nabla_p(Na_iR_m^p+Na_mR_i^p)\Big]\nb\\
           &&+2\nabla^2\nabla_p(Na_{(i}R_{j)}^p)\nb\\
           && +2g_{ij}\nabla^m\nabla^n \nabla^p(Na_{(n}R_{m)p})\Bigg\},
\eqn
%
\bqn\lb{a5}
(F_0^a)_{ij} &=&  -\frac{1}{2} g_{ij} a^k a_k +a_i a_j, \nb\\
(F_1^a)_{ij} &=&  -\frac{1}{2} g_{ij} (a_k a^k)^2 + 2 (a_k a^k) a_i a_j,\nb\\
(F_2^a)_{ij} &=&  -\frac{1}{2} g_{ij} (a_k^{\;\; k})^2 + 2 a_k^{\;\; k} a_{ij}\nb\\
             &&   - \frac{1}{N} \Big[2 \nabla_{(i} (N a_{j)} a_k^{\;\; k}) - g_{ij} \nabla^l (a_l N a_k^{\;\; k})\Big],\nb\\
(F_3^a)_{ij} &=&   -\frac{1}{2} g_{ij} (a_k a^k) a_l^{\;\; l} + a^k_{\;\;k} a_ia_j + a_k a^k a_{ij}\nb\\
             &&   - \frac{1}{N} \Big[ \nabla_{(i} (N a_{j)} a_k a^k) - \frac{1}{2} g_{ij} \nabla^l (a_l N a_ka^k)\Big],\nb\\
(F_4^a)_{ij} &=&  - \frac{1}{2} g_{ij} a^{mn} a_{mn} + 2a^k_{\;\; i} a_{kj} \nb\\
             &&   - \frac{1}{N} \Big[\nabla^k (2 N a_{(i} a_{j)k} - N a_{ij} a_k)\Big], \nb\\
(F_5^a)_{ij} &=&  -\frac{1}{2} g_{ij} (a_k a^k ) R + a_i a_j R + a^k a_k R_{ij} \nb\\
              &&  + \frac{1}{N} \Big[ g_{ij} \nabla^2 (N a_k a^k) - \nabla_i \nabla_j (N a_k a^k)\Big], \nb\\ 
(F_6^a)_{ij} &=&   -\frac{1}{2} g_{ij} a_m a_n R^{mn} +2 a^m R_{m (i} a_{j)} \nb\\
              &&  - \frac{1}{2N} \Big[ 2 \nabla^k \nabla_{(i} (a_{j)} N a_k) - \nabla^2 (N a_i a_j) \nb\\
               && - g_{ij} \nabla^m \nabla^n (N a_m a_n)\Big], \nb\\ 
(F_7^a)_{ij} &=&  -\frac{1}{2} g_{ij} R  a_k^{\;\; k} +  a_k^{\;\; k} R_{ij} + R a_{ij} \nb\\
             &&   + \frac{1}{N} \Big[ g_{ij} \nabla^2 (N  a_k^{\;\; k}) - \nabla_i \nabla_j (N  a_k^{\;\; k}) \nb\\
             &&  - \nabla_{(i} (N R a_{j)}) + \frac{1}{2} g_{ij} \nabla^k (N R a_k)\Big], \nb\\
(F_8^a)_{ij} &=&  -\frac{1}{2} g_{ij} (\Delta a_k)^2 + (\Delta a_i) (\Delta a_j) + 2 \Delta a^k \nabla_{(i} \nabla_{j)} a_k \nb\\
             &&   + \frac{1}{N} \Big[\nabla_k [a_{(i} \nabla^k (N \Delta a_{j)}) + a_{(i} \nabla_{j)} (N \Delta a^k)\nb\\
             &&   - a^k \nabla_{(i} (N \Delta a_{j)}) + g_{ij} N a^{l k} \Delta a_l  - N a_{ij} \Delta a^k ]\nb\\
             &&  -  2 \nabla_{(i} (N a_{j)k} \Delta a^k)\Big],
\eqn
%
\bqn\lb{a6}
(F_1^\varphi)_{ij} &=&     -\frac{1}{2} g_{ij} \varphi {\cal{G}}^{mn}K_{mn}\nb\\
                   &&  + \frac{1}{2\sqrt{g} N}  \partial_t (\sqrt{g} \varphi {\cal{G}}_{ij}) -
                                                                                2 \varphi K_{(i}^l R_{j) l} \nb\\
                   &&       + \frac{1}{2} \varphi (K R_{ij} + K_{ij} R - 2 K_{ij} \Lambda_g ) \nb\\
                   &&      + \frac{1}{2N} \bigg\{2 {\cal{G}}_{k(i} \nabla^k (N_{j)} \varphi)-{\cal{G}}_{ij} \nabla^k (\varphi N_k) \nb\\
                   &&      +  g_{ij}\nabla^2 (N \varphi K) - \nabla_i \nabla_j (N \varphi K) \nb\\
                   &&+ 2  \nabla^k \nabla_{(i} (K_{j)k} \varphi N),\nb\\
                   &&      - \nabla^2 (N \varphi K_{ij}) - g_{ij} \nabla^k \nabla^l (N \varphi K_{kl})\bigg\}, \nb\\
(F_2^\varphi)_{ij} &=&   - \frac{1}{2}g_{ij} \varphi {\cal{G}}^{mn} \nabla_m\nabla_n \varphi \nb\\
                   &&     - 2 \varphi \nabla_{(i} \nabla^k R_{j)k} + \frac{1}{2} \varphi (R- 2 \Lambda_g) \nabla_i \nabla_j \varphi \nb\\
                   &&     - \frac{1}{N} \bigg\{- \frac{1}{2} (R_{ij} + g_{ij} \nabla^2 - \nabla_i \nabla_j )(N \varphi \nabla^2 \varphi)\nb\\
                   &&      - \nabla_k \nabla_{(i} (N \varphi \nabla^k \nabla_{j)} \varphi) + \frac{1}{2}\nabla^2 (N \varphi \nabla_i \nabla_j \varphi) \nb\\
                   &&      + \frac{g_{ij}}{2} \nabla^k \nabla^l ( N \varphi \nabla_k \nabla_l \varphi)\nb\\
                   &&      - {\cal{G}}_{k (i} \nabla^k (N \varphi \nabla_{j)}\varphi ) + \frac{1}{2} {\cal{G}}_{ij} \nabla^k (N \varphi \nabla_k \varphi)\bigg\},\nb\\
(F_3^\varphi)_{ij} &=&    - \frac{1}{2}g_{ij} \varphi {\cal{G}}^{mn} a_m\nabla_n \varphi \nb\\
                   &&    -\varphi ( a_{(i} R_{j)k} \nabla^k \varphi + a^k R_{k(i} \nabla_{j)} \varphi)\nb\\
                   &&      + \frac{1}{2} (R - 2 \Lambda_g)  \varphi a_{(i} \nabla_{j)} \varphi \nb\\
                   &&      - \frac{1}{N}\bigg\{ - \frac{1}{2} (R_{ij} + g_{ij} \nabla^2 - \nabla_i \nabla_j ) (N \varphi a^k \nabla_k \varphi)\nb\\
                   &&     - \frac{1}{2} \nabla^k \Big[  \nabla_{(i} (\nabla_{j)} \varphi N \varphi)+ \nabla_{(i} (a_{j)} \varphi N \nabla_k \varphi) \Big] \nb\\
                   &&     + \frac{1}{2}\nabla^2 (N \varphi a_{(i} \nabla_{j)} \varphi) \nb\\
                   &&+ \frac{g_{ij}}{2} \nabla^k \nabla^l (N \varphi a_k \nabla_l \varphi)\bigg\}, \nb\\ 
(F_4^\varphi)_{ij} &=&    - \frac{1}{2}g_{ij} \hat{{\cal{G}}}^{mnkl}K_{mn} a_{(k}\nabla_{l)}\varphi \nb\\
                   &&    + \frac{1}{2 \sqrt{g} N} \partial_t [\sqrt{g} {\cal{G}}_{ij}^{\;\;k l} a_{(l} \nabla_{k)} \varphi] \nb\\
                   &&      + \frac{1}{2N} \nabla^l \Big[a_l N_{(i} \nabla_{j)} \varphi +  N_{(i} a_{j)} \nabla_l  \varphi \nb\\
                   &&      -  N_l a_{(i} \nabla_{j)} \varphi + 2 g_{ij} N_l a^k \nabla_k \varphi \Big] \nb\\
                   &&      + \frac{1}{N} \nabla_{(i} (N N_{j)} a^k \nabla_k \varphi)\nb\\
                   &&      + a^k K_{k(i} \nabla_{j)} \varphi + a_{(i} K_{j)k} \nabla^k \varphi \nb\\
                   &&      - K a_{(i} \nabla_{j)} \varphi - K_{ij} a^k \nabla_k \varphi, \nb\\
(F_5^\varphi)_{ij} &=&     -\frac{1}{2} g_{ij} \hat{{\cal{G}}}^{mnkl}[a_{(k}\nabla_{l)}\varphi][\nabla_m\nabla_n\varphi]\nb\\
                    && -a_{(i} \nabla^k \nabla_{j)} \varphi \nabla_k \varphi - a_k \nabla^k \nabla_{(i} \varphi \nabla_{j)}\varphi\nb\\
                   &&     + a_{(i} \nabla_{j)} \varphi \nabla^2 \varphi + a^k \nabla_k \varphi \nabla_i\nabla_j \varphi \nb\\
                   &&     + \frac{1}{2N} \bigg\{ \nabla^k (N \varphi a_k \nabla_i \varphi \nabla_j \varphi) \nb\\
                   &&   - 2 \nabla_{(i} (N \nabla_{j)} \varphi a^k
                          \nabla_k \varphi) \nb\\
                   &&     +g_{ij} \nabla^l (\nabla_l \varphi a^k \nabla_k \varphi)\bigg\}, \nb\\
(F_6^\varphi)_{ij} &=&   - \frac{1}{2} g_{ij}\hat{{\cal{G}}}^{mnkl} [a_{(m}\nabla_{n)}\varphi][a_{(k}\nabla_{l)}\varphi]\nb\\
                     &&-\frac{1}{2} (a^k \nabla_i \varphi- a_i \nabla^k \varphi) (a_k \nabla_j \varphi - a_j \nabla_k \varphi), \nb\\
(F_7^\varphi)_{ij} &=&   -\frac{1}{2} g_{ij}   \hat{{\cal{G}}}^{mnkl} [\nabla_{(n}\varphi][a_{m)(k}][\nabla_{l)}\varphi]\nb\\
                    && -\frac{1}{2} a_k^{\;\; k} \nabla_i \varphi \nabla_j \varphi - \frac{1}{2} a_{ij} \nabla^k \varphi \nabla_k \varphi \nb\\
                     &&     +  a^k_{(i} \nabla_{j)} \varphi \nabla_k \varphi - \frac{1}{2N}\bigg \{- \nabla_{(i} (N a_{j)} \nabla_k \varphi \nabla^k \varphi) \nb\\ &&+ \nabla^k (N a_{(i} \nabla_{j)} \varphi \nabla_k \varphi)\nb\\
                   &&      + \frac{g_{ij}}{2} \nabla^k (N a_k \nabla^m \varphi \nabla_m \varphi) \nb\\
                   &&- \frac{1}{2}\nabla^k (N a_k \nabla_i \varphi \nabla_j \varphi)\bigg\}, \nb\\
(F_8^\varphi)_{ij} &=&    - \frac{1}{2} g_{ij} (\nabla^2 \varphi+a_k\nabla^k\varphi)^2\nb\\
                   &&    -2 (\nabla^2 \varphi + a_k \nabla^k \varphi) (\nabla_i \nabla_j \varphi + a_i \nabla_j \varphi ) \nb\\
                   &&      -\frac{1}{N} \bigg \{ - 2 \nabla_{(j} [N \nabla_{i)} \varphi (\nabla^2 \varphi + a_k \nabla^k \varphi)] \nb\\
                   &&       + g_{ij} \nabla^l [N (\nabla^2 \varphi + a_k \nabla^k \varphi) \nabla_l \varphi]\bigg\}, \nb\\
(F_9^\varphi)_{ij} &=&    - \frac{1}{2} g_{ij}(\nabla^2 \varphi+a_k\nabla^k\varphi)K \nb\\
                    && - (\nabla^2 \varphi + a_k \nabla^k \varphi) K_{ij} \nb\\
                   &&     - (\nabla_i \nabla_j \varphi + a_i \nabla_j \varphi ) K \nb\\
                  &&     +\frac{1}{2 \sqrt{g} N}  \partial_t [\sqrt{g} (\nabla^2 \varphi + a_k \nabla^k \varphi) g_{ij}] \nb\\
                    &&      - \frac{1}{N}\bigg\{ - \nabla_{(j} [ N_{i)} (\nabla^2 \varphi + a_k \nabla^k \varphi)] \nb\\
                    &&     + \frac{1}{2} g_{ij} \nabla^l [ N_l (\nabla^2 \varphi + a_k \nabla^k \varphi) ]\nb\\
                    &&    -  \nabla_{(j} (N K \nabla_{i)} \varphi) + \frac{1}{2} g_{ij} \nabla_k (N K \nabla^k \varphi)
                            \bigg\}.
\eqn

\section*{Appendix C: Scalar-tensor extension}
\renewcommand{\theequation}{C.\arabic{equation}} \setcounter{equation}{0}

For simplicity let us consider the case with $z=3$ in the UV and
introduce extra scalar fields $\chi_{\alpha}$
($\alpha=1,2,\cdots,n$). Since $z=3$ in the UV, $\chi_{\alpha}$
naturally have vanishing scaling dimension in the UV. This means that
replacing a constant in the action by a function of $\chi_{\alpha}$ does
not spoil the power-counting renormalizability of the theory. (A similar
statement does not apply to $\sigma$,  since $\sigma$ has the scaling
dimension $(2z-2)=4\ne 0$ in the UV.) Furthermore, let us suppose that
$\chi_{\alpha}$ couple to matter universally. To be more precise, we
suppose that the  variables 
$$
\left(\tilde{N}, \tilde{N}^i,
\tilde{g}_{ij}\right),
$$
 couple to matter universally as if they were the
lapse, the shift and the spatial metric, where
\begin{eqnarray}
 \tilde{N} & \equiv & F\, N,\nonumber\\
 \tilde{N}^i & \equiv &  N^i + Ng^{ij}\partial_j\varphi, \nonumber\\
 \tilde{g}_{ij} & \equiv & \Omega^2\, g_{ij},
  \label{eqn:new-metric-variables}
\end{eqnarray}
where $F$ and $\Omega$ are non-vanishing functions of the scalar fields
$\chi_{\alpha}$. It is easy to see that these new variables are
invariant under the $U(1)$ transformation and transform under the
infinitesimal foliation preserving diffeomorphism as
\begin{eqnarray}
 \delta \tilde{N} & = & \partial_t(\tilde{N} f)
  + \xi^i\partial_i\tilde{N},   \nonumber\\
 \delta \tilde{N}^i & = & \partial_t (\tilde{N}^i f) + \partial_t \xi^i
  + {\cal L}_{\xi}\tilde{N}^i, \nonumber\\
 \delta (\tilde{N}_i) & = & \partial_t (\tilde{N}_i f)
  + \tilde{g}_{ij}\partial_t \xi^j
  + {\cal L}_{\xi}\tilde{N}_i, \nonumber\\
 \delta \tilde{g}_{ij} & = & f\partial_t \tilde{g}_{ij} + {\cal L}_{\xi}\tilde{g}_{ij},
\end{eqnarray}
where
\begin{equation}
 \tilde{N}_i \equiv \tilde{g}_{ij}\tilde{N}^j
  = \Omega^2\, (N_i+N\partial_i\varphi),
\end{equation}
and $\tilde{g}^{ij}=\Omega^{-2}g^{\ij}$ is the inverse of
$\tilde{g}_{ij}$.

For simplicity, from now on, let us consider the case with just one
extra scalar field (we shall denote it as $\chi$). As already stated
above, since $z=3$ is assumed in the UV, $\chi$ naturally has vanishing
scaling dimension in the UV. Since the scalar $\sigma$ defined in
(\ref{2.6}) has the scaling dimension $(2z-2)=4$ in the UV,
the renormalizable potential for $\chi$ and $\sigma$ is in general of
the form
\begin{equation}
 V(\chi) + \sigma U(\chi),
\end{equation}
where $V(\chi)$ and $U(\chi)$ are general functions of $\chi$. Let us
suppose that $V(\chi)$ has a minimum at $\chi=\chi_0$ and Taylor expand
the two functions as
\begin{eqnarray}
 V(\chi) & \simeq & V_0 + \frac{1}{2}m^2(\chi-\chi_0)^2, \nonumber\\
 U(\chi) & \simeq & U_0 + U_1(\chi-\chi_0),
\end{eqnarray}
where $V_0$, $m$ and $U_{0,1}$ are constants. We also Taylor expand the
functions $F$ and $\Omega$ in (\ref{eqn:new-metric-variables}) as
\begin{eqnarray}
 F(\chi) & \simeq & F_0 + F_1(\chi-\chi_0), \nonumber\\
 \Omega(\chi) & \simeq & \Omega_0 + \Omega_1(\chi-\chi_0),
\end{eqnarray}
where $F_{0,1}$ and $\Omega_{0,1}$ are constants.

At energies sufficiently lower than $m$, the extra scalar $\chi$ can be
integrated out (by setting $\chi-\chi_0\simeq -\sigma U_1/m^2$) and
we obtain
\begin{eqnarray}
 \frac{F}{F_0} & \simeq &
  1 - \frac{F_1U_1}{F_0m^2}\sigma, \nonumber\\
 \frac{\Omega}{\Omega_0} & \simeq &
  1 - \frac{\Omega_1U_1}{\Omega_0m^2}\sigma.
\end{eqnarray}
Note that $F_0$ and $\Omega_0$ has been set to unity by redefinition of
the units of time and spatial coordinates. For this reason the
parameters $\gamma_1$ (in the non-projectable case) and $g_1$ (in the
projectable case) can no longer be rescaled and thus their values have
physical meaning.

\section*{Appendix D: The functions  appearing in $h_{\mu\nu}$ and their main properties}
\renewcommand{\theequation}{D.\arabic{equation}} \setcounter{equation}{0}

The functions appearing in $h_{\mu\nu}$ of Eq.(\ref{4.20}) are defined as,
  \bqn
 \lb{4.21}
  U &\equiv& \int\frac{\rho(\mathbf{x}',t)}{|\mathbf{x}-\mathbf{x}'|}d^3x',  \nb\\
  \chi &\equiv& -\int\rho(\mathbf{x}',t)|\mathbf{x}-\mathbf{x}'|d^3x',  \nb\\
  V_j &\equiv& \int\frac{\rho(\mathbf{x}',t)v'_j}{|\mathbf{x}-\mathbf{x}'|}d^3x',  \nb\\
 W_j &\equiv& \int\frac{\rho(\mathbf{x}',t)\mathbf{v}' \cdot (\mathbf{x}-\mathbf{x}')(x-x')_j}{|\mathbf{x}-\mathbf{x}'|^3}d^3x',  \nb\\
 U_{ij} &\equiv& \int\frac{\rho(\mathbf{x}',t)(x-x')_i(x-x')_j}{|\mathbf{x}-\mathbf{x}'|^3}d^3x',  \nb\\
 \Phi_W &\equiv& \int\rho'\rho''\frac{\mathbf{x}-\mathbf{x}'}{|\mathbf{x}-\mathbf{x}'|^3}\nb\\
 && \times\left(\frac{\mathbf{x}'-\mathbf{x}''}{|\mathbf{x}-\mathbf{x}''|}-\frac{\mathbf{x}-\mathbf{x}''}{|\mathbf{x}'-\mathbf{x}''|}\right)d^3x'd^3x'',  \nb\\
 \Phi_1 &\equiv& \int\frac{\rho' v'^2}{|\mathbf{x}-\mathbf{x}'|}d^3x',  \nb\\
   \Phi_2 &\equiv& \int\frac{\rho'U'}{|\mathbf{x}-\mathbf{x}'|}d^3x',  \nb\\
    \Phi_3 &\equiv& \int\frac{\rho'\Pi'}{|\mathbf{x}-\mathbf{x}'|}d^3x',  \nb\\
     \Phi_4 &\equiv& \int\frac{p'}{|\mathbf{x}-\mathbf{x}'|}d^3x',  \nb\\
 \mathfrak{A} &\equiv& \int\frac{\rho'[\mathbf{v}'\cdot(\mathbf{x}-\mathbf{x}')]^2}{|\mathbf{x}-\mathbf{x}'|^3}d^3x',  \nb\\
  \mathfrak{B} &\equiv& \int\frac{\rho'}{|\mathbf{x}-\mathbf{x}'|}(\mathbf{x}-\mathbf{x}')\cdot \frac{d\mathbf{v}'}{dt} d^3x'.
 \eqn
Then, from the continuity equation (\ref{4.19}), we find that
\bqn
\lb{4.24}
&& \partial^2 U = -4\pi G_{N} \rho,~~~\nabla^2\chi=-2U,~~~\chi_{,0j}=V_j-W_j,\nb\\
&& \partial^2 V_j = -4\pi G_{N} \rho v_j, ~~~V_{j,j}=-U_{,0},\nb\\
&& \partial^2 \Phi_1 = -4\pi G_{N} \rho v^2,~~~\partial^2 \Phi_2=-4\pi G_{N} \rho U=U \partial^2 U,\nb\\
&& \partial^2 \Phi_3 = -4\pi G_{N} \rho \Pi,~~~\partial^2 \Phi_4=-4\pi G_{N} p,\nb\\
&& 2\chi_{,ij}U_{,ij} = \partial^2 (\Phi_W+2U^2-3\Phi_2),\nb\\
&& \chi_{,00} = \mathfrak{A} + \mathfrak{B}  -\Phi_1,
\eqn
where $\Pi$ denotes the specific energy density and is equal to the ratio of energy density to the rest-mass density \cite{Will}.

\section*{Appendix E: Hamiltonian constraint and the trace part of the dynamical equations to
${\cal{O}}(4)$}
\renewcommand{\theequation}{E.\arabic{equation}} \setcounter{equation}{0}

The Hamiltonian constraint and the trace part of the dynamical equations to
${\cal{O}}(4)$ are given,  respectively,  by
 \bqn
 \lb{D1a}
 8\pi J^t&=&\gamma_1R+\beta_0(2a_i^i+a_ia^i)\nb\\
 &&+2\frac{\sigma_1}{N}\nabla_i(Aa^i)-\frac{\sigma_2}{N}\nabla^2A,\\
 \lb{D1b}
 8\pi g^{ij}\tau_{ij}&=&\frac{g^k_ig_{jk}}{N\sqrt{g}}\frac{\partial}{\partial
 t}\left[\sqrt{g}\left(-K^{ij}+\lambda
 g^{ij}K\right)\right]+\frac{AR}{2N}\nb\\
 &&-\frac{\beta_0}{2}a^ka_k-\frac{2}{N}\nabla^2A+\frac{\gamma_1}{2}R-\frac{2\gamma_1}{N}\nabla^2N\nb\\
 &&+\frac{\sigma_2}{2}f(1-a_1f)\left(\partial^2\Phi_2-\frac{1}{2}\partial^2U^2\right).
 \eqn
Combining  Eq.(\ref{4.17b}) with Eqs.(\ref{D1a}) and (\ref{D1b}) so that  the
$R$ terms can be canceled, we obtain
 \bqn
 \lb{D2a}
&& (2\beta_0+\sigma_2\gamma_1)a_i^i+(\beta_0+\sigma_1\gamma_1)a_ia^i\nb\\
&& ~~~~ +2\frac{\sigma_1}{N}\nabla_i(Aa^i)-\frac{\sigma_2}{N}\nabla^2A = 8\pi (J^t-\gamma_1J_A),  ~~~~~~~\\
 \lb{D2b}
&& -(1-3\lambda)K_{,0}-2f(\gamma+a_2f)\partial^2\Phi_2\nb\\
&&+\frac{\sigma_2}{2}f(1-a_1f)\left(\partial^2\Phi_2-\frac{1}{2}\partial^2U^2\right)\nb\\
&& -\frac{a_ka^k}{2}\left(\beta_0-\gamma_1\sigma_1\right) +\frac{\gamma_1}{2}\sigma_2a^k_k \nb\\
&& -\frac{2}{N}\nabla^2A-\frac{2\gamma_1}{N}\nabla^2N = 8\pi
\left(g^{ij}\tau_{ij}-\frac{\gamma_1}{2}J_A\right).
 \eqn
Finally, from Eqs.(\ref{D2a}) and (\ref{D2b})  we can   cancel the $\nabla^2A$ terms to get an equation, which contains only
$h_{00}$ terms in the fourth-order of $v$. Solving this equation we find Eq.(\ref{4.37}). The explicit expression of this equation
 is too complicated to be written down here.

The coefficients $h_{n}$ in Eq.(\ref{4.37}) are given by,
 \bqn
 \lb{4.37a}
h_{1}&=&-2E_0^{-1}(1-3\lambda)(c+2d+3\gamma+3a_2f)\nb\\
&&\times(2a_1\beta_0-\sigma_2+a_1\gamma_1\sigma_2), \nb\\
h_{2} &=&E_0^{-1}\Big\{-16 a_1^2 \gamma _1 \varkappa  \left(\beta
_0+2 \gamma _1\right)\nb\\
&&+2 a_1 \Big[2 \beta _0 \Big(a_2 \left(4 \gamma _1 \varkappa +f
(3-9
   \lambda )\right)\nb\\
   &&-3 \lambda  (c+3 \gamma +2 d)+c+3 \gamma +2 d+4 \varkappa \Big)\nb\\
   &&+\gamma _1 \big[\sigma _2 \big(3 a_2 f (1-3 \lambda
   )-3 \lambda  (c+3 \gamma +2 d)\nb\\
   &&+c+3 \gamma +2 d+12 \varkappa \big)+16 \varkappa  \left(a_2 \gamma _1-2\right)\big]\Big]\nb\\
   &&-2 \sigma
   _2 \Big[a_2 \left(4 \gamma _1 \varkappa +f (3-9 \lambda )\right)-3 \lambda  (c+3 \gamma +2 d)\nb\\
   &&+c+3 \gamma +2 d+4 \varkappa \Big]+32
   \varkappa  \left(a_2 \gamma _1-1\right)\Big\}, \nb\\
h_{3}&=&2E_0^{-1}\Big\{-2 \gamma _1 \sigma _1 \left(a_1
f-1\right){}^2 \left[a_1 \left(\beta _0+2 \gamma
_1\right)+2\right]\nb\\
&&+\sigma _2 \Big[\beta _0 \left(a_1
   f-1\right) \left(a_1 \left(2 \gamma _1 \left(a_1 f-1\right)-3 f\right)+1\right)\nb\\
   &&+\gamma _1 \big[4 \big(a_1 \left(\gamma _1 \left(a_1
   f-1\right){}^2+f \left(a_1 f-\gamma \right)\right)+6 \gamma \nb\\
   &&-1\big)+\sigma _1 \left(a_1 f-1\right){}^2\big]+4 \gamma  f\Big]\nb\\
   &&+a_2
   f \Big[48 \varkappa  \left(a_1 \gamma _1+1\right)^2+8 \beta _0 (3 a_1^2 \gamma _1 \varkappa -a_1 f\nb\\
   &&+3)+4 \sigma _2
   \left(f-\gamma _1 \left(a_1 (f+6 \varkappa )-6\right)\right)-3 \gamma _1 \sigma _2^2\Big]\nb\\
   &&+2 \beta _0 \Big[a_1 \big(a_1 f
   \big(\beta _0 \left(a_1 f-2\right)+2 (\gamma _1 \left(a_1 f-2\right)\nb\\
   &&+f)\big)+\beta _0+2 \gamma _1-4 \gamma  f\big)+12
   \gamma -2\Big]\nb\\
   &&-\sigma _2^2 \left(a_1 f \left(2 \gamma _1 \left(a_1 f-1\right)-f\right)+3 \gamma  \gamma _1+f\right)\Big\}, \nb\\
h_{4}&=&-\varkappa\frac{16}{E_0}\left[2+a_1\gamma_1(4-\sigma_2+a_1\beta_0+2a_1\gamma_1)\right], \nb\\
h_{5}
&=&\varkappa\frac{24}{E_0}[2a_1\beta_0(1+a_2\gamma_1)+a_1\gamma_1
(4a_2\gamma_1+\sigma_2)\nb\\
&&-a_2\gamma_1(\sigma_2-4)-\sigma_2], \nb\\
h_{6} &=&E_0^{-1}\Big\{\sigma _2 \Big[\sigma _2 \Big(-a_1 f^2+\gamma
_1 \big(f \left(a_1 \left(2 a_1 f-3\right)+a_2\right)\nb\\
&&+\gamma +2\big)+f\Big)-4 a_1
   \gamma _1^2 \left(a_1 f-1\right)^2\nb\\
   &&-4 \gamma _1 \left(a_1 f \left(a_1 f-2\right)+2 a_2 f+2 \gamma +3\right)\Big]\nb\\
   &&+\beta _0 \Big[2
   a_1 \gamma _1 \sigma _1 \left(a_1 f-1\right)^2\nb\\
   &&-\sigma _2 \left(a_1 f-1\right) \big(a_1 \left(2 \gamma _1 \left(a_1 f-1\right)-3
   f\right)+1\big)\nb\\
   &&-4 a_1 \gamma _1 \left(a_1 f-1\right)^2-4 \big(a_1 f \left(a_1 f-2\right)+2 a_2 f\nb\\
   &&+2 \gamma +3\big)\Big]-2 a_1
   \beta _0^2 \left(a_1 f-1\right)^2\nb\\
   &&+\sigma _1 \left(a_1 f-1\right) \Big[\gamma _1 \Big(a_1 \big(4 \gamma _1 \left(a_1 f-1\right)\nb\\
   &&-3 f \left(\sigma _2-4\right)\big)+\sigma _2-4\Big)+8 f\Big]\Big\},
 \eqn
where
 \bqn
 \lb{4.38}
E_0&=&8\beta_0+(8-\sigma_2)\gamma_1\sigma_2.
 \eqn

\section*{Appendix F: Independent PPN parameters for $N = N(t, x)$}
\renewcommand{\theequation}{F.\arabic{equation}} \setcounter{equation}{0}

 Combining Eqs.(\ref{4.32}), (\ref{4.35}), (\ref{4.36a}) and (\ref{4.38a})-(\ref{4.38h}),  finally we find that  the eleven PPN parameters are given by
\bqn \lb{4.39} \beta&=&\frac{\left[a_1 \gamma _1 \left(\sigma
_2-4\right)-4\right]^{-3}}{4\varkappa}\Big\{\gamma _1 \sigma _2^3
\big[a_1^2 \gamma _1 \varkappa  \left(2 a_1 \gamma
_1-1\right)\nb\\
&&+1\big]-4 \gamma _1 \sigma _2^2 \big[a_1 \varkappa
    \big(a_1 \gamma _1 \left(2 a_1 \gamma _1+1\right) \left(a_1 \varkappa +3\right)\nb\\
    &&-5\big)+1\big]-8 \sigma _2 \big[\gamma _1 \sigma
   _1 \left(a_1^2 \gamma _1 \varkappa +1\right)^2\nb\\
   &&-2 \varkappa  \left(a_1 \gamma _1+1\right){}^2 \left(a_1 \gamma _1 \left(5 a_1
   \varkappa +6\right)-1\right)\big]\nb\\
   &&-32 \gamma _1 \sigma _1 \left(a_1 \varkappa -1\right) \left(a_1^2 \gamma _1 \varkappa
   +1\right)^2\nb\\
   &&+64 \varkappa  \left(a_1 \gamma _1+1\right)^2 \big[a_1 [\gamma _1 \left(a_1 \varkappa  \left(a_1 \varkappa
   -3\right)-2\right)\nb\\
   &&-\varkappa ]-3\big]\Big\},\nb\\
\gamma&=&-\frac{a_2 \gamma _1 \left(4 a_1 \varkappa +\sigma
_2-4\right)+4 a_1^2 \gamma _1 \varkappa +4 a_1 \varkappa +\sigma
_2}{a_1 \gamma _1
   \left(\sigma _2-4\right)-4},\nb\\
\alpha_1&=&\frac{16 \left(a_1-a_2\right) \gamma _1+16}{a_1 \gamma _1
\left(\sigma _2-4\right)-4}\nb\\
&&+\frac{4 \sigma _2 \left(\gamma _1 \left(a_1 (2
   \varkappa -1)+a_2\right)+1\right)}{a_1 \gamma _1 \left(\sigma _2-4\right)-4}\nb\\
   &&+\frac{16 \varkappa  \left(\left(a_1+a_2-2\right) a_1
   \gamma _1+a_1-2\right)}{a_1 \gamma _1 \left(\sigma _2-4\right)-4},\nb\\
\alpha_2&=&(\lambda -1)^{-1} \varkappa^{-1}  \left[a_1 \gamma _1 \left(\sigma _2-4\right)-4\right]^{-2}\nb\\
&&\times\Big\{8 \sigma _2 \varkappa  \left(a_1 \gamma _1+1\right)
\big[a_1 \lambda  [\left(3 a_1-4\right) \gamma _1 \varkappa +\gamma
_1+3]\nb\\
&&-a_1
   \left(\left(a_1-2\right) \gamma _1 \varkappa +\gamma _1+1\right)-3 \lambda +1\big]\nb\\
   &&+\sigma _2^2 \big[a_1 \gamma _1 \varkappa  \big(a_1
   \gamma _1 (-\lambda +(4 \lambda -2) \varkappa +1)\nb\\
   &&+6 \lambda -2\big)+3 \lambda -1\big]\nb\\
   &&+16 \varkappa  \left(a_1 \gamma _1+1\right)^2
   \big[\varkappa  \big(\left(a_1-2\right) a_1 (3 \lambda -1)+4 \lambda \nb\\
   &&-2\big)-\lambda +1\big]\Big\},\nb\\
\zeta_1&=&-\zeta_B=\frac{(3 \lambda -1) \left[4 a_1 \varkappa
\left(a_1 \gamma _1+1\right)+\sigma _2\right]}{(\lambda -1)
\varkappa \left[a_1 \gamma _1
   \left(\sigma _2-4\right)-4\right]^2}\nb\\
   &&\times[a_1 \gamma _1 \sigma _2 \varkappa +4 \left(a_1-1\right) \varkappa  \left(a_1 \gamma _1+1\right)+\sigma _2],\nb\\
\alpha_3&=& \xi  =  \zeta_2 = \zeta_3 =  \zeta_4= 0,
 \eqn
and
   \bqn
    \lb{4.40}
 \beta_0 &=&\frac{8 \varkappa  \left[a_1 \gamma _1 \sigma _2-2 \left(a_1 \gamma _1+1\right)^2\right]+\gamma _1 \left(\sigma _2-8\right) \sigma _2}{8
   a_1^2 \gamma _1 \varkappa +8},\nb\\
 d &=&\frac{2 \varkappa  \left[2 a_1^2 \gamma _1 (3 \lambda -1)-8 \lambda +4\right]+(3 \lambda -1) \sigma
   _2}{2 (\lambda -1) \left[a_1 \gamma _1 \left(\sigma
   _2-4\right)-4\right]}\nb\\
   &&+\frac{a_1 \varkappa  \left[\gamma _1 (2 \lambda -1) \left(\sigma _2-4\right)+6 \lambda -2\right]}{(\lambda -1) \left[a_1 \gamma _1
   \left(\sigma _2-4\right)-4\right]}.
 \eqn


\end{document}